\documentclass[onecolumn,prx,superscriptaddress,longbibliography,nofootinbib]{revtex4-2}

\usepackage[dvips]{graphicx} % for figures
\usepackage{amsfonts,amscd,amsmath,amsthm}
\usepackage{mathrsfs}
\usepackage{enumerate}
\usepackage{epsfig}
\usepackage{subfigure}
\usepackage{xcolor}
\usepackage[colorlinks = true]{hyperref}
\usepackage{physics}
\usepackage{epstopdf}
\usepackage{framed}
\usepackage{multirow}
\usepackage{color}
\usepackage{longtable}
\usepackage{comment}
\usepackage[ruled,vlined]{algorithm2e}
\usepackage[most]{tcolorbox}
\usepackage{enumitem}

\graphicspath{{./figure/}}

\usepackage{tikz}
\usetikzlibrary{tikzmark, calc, fit, positioning}
\usetikzlibrary{shapes}

\newtheorem{theorem}{Theorem}
\newtheorem{lemma}{Lemma}

\newtheorem{definition}{Definition}
\newtheorem{proposition}{Proposition}

\newtcolorbox[auto counter]{mybox}[2][]{
	enhanced,
	breakable,
	colback=blue!5!white,
	colframe=blue!75!black,
	fonttitle=\bfseries,
	title=Box \thetcbcounter: #2,#1
}

\begin{document}

\title{Measurement-device-independent resource characterization protocols}

\author{Chenxu Li}
\affiliation{Center for Quantum Information, Institute for Interdisciplinary Information Sciences, Tsinghua University, Beijing 100084, China}
\author{Mingze Xu}
\affiliation{Center for Quantum Information, Institute for Interdisciplinary Information Sciences, Tsinghua University, Beijing 100084, China}
\affiliation{Zhili College, Tsinghua University, Beijing 100084, China}
\author{Hao Dai}
\affiliation{Beijing Institute of Mathematical Sciences and Applications (BIMSA), Huairou District,
Beijing 101408, China}
\author{Xiongfeng Ma}
\email{xma@tsinghua.edu.cn}
\affiliation{Center for Quantum Information, Institute for Interdisciplinary Information Sciences, Tsinghua University, Beijing 100084, China}

\begin{abstract}
Measurement-device-independent (MDI) quantum information processing tasks are important subroutines in quantum information science because they are robust against any type of measurement imperfections. In this work, we propose a framework of MDI resource characterization protocols that unifies and generalizes these tasks. We show that resources that do not increase under local operation and shared randomness can be characterized with any untrusted measurement, and we provide a general procedure to convert such resource characterization tasks into MDI protocols. We then focus on applying our framework to two cases that satisfy the criteria: the resource theory of bipartite and multipartite entanglement, and the resource theory of quantum memories. We demonstrate several MDI characterization protocols for these resources. These protocols are either novel or generalize existing ones from the literature. We also show that MDI quantum key distribution can be viewed as an MDI quantification protocol for quantum memory. 

\end{abstract}

\maketitle

\tableofcontents

\section{Introduction}

One of the main goals of quantum information science is to study the quantitative relationship between the properties of quantum features beyond classical theories and the advantages they bring to information processing tasks. Quantum resource theories \cite{QRTReview,GeneralResource} provide valuable insights into this approach. In a quantum resource theory, resourceful objects are those that cannot be created for free, i.e., they can only be converted from another resource object by free operations, or created by a resourceful operation. For instance, bipartite entanglement is the key resource in quantum communication tasks \cite{QKDReview}. General resource theories are much more versatile. A resource theory can be static, where the involved objects are quantum states, or dynamical, where the objects are quantum channels \cite{liu2019resourcetheoriesquantumchannels,YuanChannelResource}. For example, in the resource theory of quantum memories \cite{QRTMemories}, free objects are entanglement-breaking (EB) channels \cite{EBChannels}. As a resource, channel quantumness can be seen as a dynamical version of bipartite quantum entanglement.

Given the importance of quantum resources, their characterization becomes a crucial task for practical experiments and applications. Informally, after characterizing a resource, we can infer its existence, estimate its amount, and benchmark its performance in desired quantum information processing tasks. In some cases, characterization only requires us to verify the existence of the resource, while in other circumstances, a quantitative evaluation of the amount of the resource is needed. We refer to these two cases as resource verification and resource quantification.

Many methods exist for resource characterization. One can conduct full state \cite{QubitTomo} or process tomography \cite{ProcessTomo} and calculate the amount of resource on a classical computer. An alternative method is to construct resource witnesses. By measuring the expectation value of a carefully designed observable, various quantum resources can be verified, such as quantum entanglement \cite{EntReview}, coherence \cite{QRTCoherence}, thermodynamics \cite{QRTThermal}, etc. Quantitative statements can also be given from the expectation values of witness operators \cite{Witness}. Additionally, if quantum memories are allowed, multiple identical copies of a quantum state can exist at the same time, allowing for the measurement of expectation values of multicopy observables. These can be used to evaluate nonlinear properties \cite{Nonlinear1,Nonlinear2}, such as entropy \cite{QEntropies}, concurrence \cite{Concurrence}, and higher-order moments \cite{Moments}. These nonlinear properties can help design new resource characterization protocols, which may offer remarkable advantages in terms of detection power compared to protocols based on single-copy operators \cite{SitanChen}.

Nevertheless, these methods require precise measurements of the expectation values of given observables, whether these observables are (generalized) Pauli strings \cite{QubitTomo,QuditTomo}, witness operators \cite{Witness}, permutations acting on multiple state copies \cite{Moments}, or orthogonal projections that yield the probabilities of measurement outcomes. This requirement implicitly demands that all devices involved be trusted, imposing security assumptions that are hard to fulfill in practice. For example, by generating forged output data, an untrusted apparatus, which can be assumed to be controlled by a malicious adversary, can potentially deceive us into regarding a free object as resourceful \cite{TimeShift,MDIEWExperiment}. The quest to eliminate dependence on devices motivates research into self-testing \cite{SelfTestingReview} and device-independent quantum information processing protocols \cite{DIQKD,DIQRNG}. 

However, a fully device-independent protocol is generally inefficient and very demanding to implement in practice. Even from a purely theoretical perspective, device-independent protocols have their drawbacks. For example, a device-independent verification of entanglement is equivalent to a loophole-free Bell test \cite{BellTest}, but not all entangled states exhibit a violation of Bell inequality \cite{Werner1989}. Therefore, a device-independent entanglement verification protocol will inevitably suffer from this fundamental limitation, as there will always exist some states that cannot be witnessed as entangled.

We then seek to trust the components that are more reliable. In realistic quantum experiments, it is more reasonable to trust the state preparation apparatus than the measurement devices. The prepared states can be experimentally verified by a trusted party in a highly protected environment, while the measurement devices do not enjoy such properties. Protocols with this level of security are called measurement-device-independent (MDI) protocols, with their first application in quantum key distribution (QKD) tasks \cite{MDIQKD}, which remove all detector side channels while remaining experimentally practical. 

Following this line of thought, Branciard et al. \cite{Branciard2013} developed an MDI entanglement witness protocol such that every entangled state can be witnessed in this way. This approach can also be interpreted as a nonlocal game with trusted quantum inputs \cite{Buscemi2012}. Many MDI protocols related to the characterization of quantum resources have been proposed, including the detection of multipartite entanglement structures \cite{Zhao2016}, estimation of the lower bound of entanglement \cite{MDIEntMeasures,PracticalQuant}, and testing of quantum memories \cite{QRTMemories}. Some of these approaches have also been demonstrated experimentally \cite{MDIEWExperiment,MDIEWExperiment2,MDIEWExperiment3,MDIChannelExperiment,MDIChannelExperiment2,MDIChannelExperiment3}. Notably, in \cite{MDIEWExperiment}, the constructed entanglement witness is shown to be secure against a time-shift attack \cite{TimeShift} targeting the measurement apparatus, while traditional entanglement witnesses malfunction under the same attack, incorrectly treating a separable state as entangled.

However, a general framework to understand MDI resource characterization protocols is still missing. It remains unclear what kinds of resources can be characterized without trusting the measurements, and there is currently no systematic way to construct MDI protocols. Additionally, it is unknown whether there is a connection between MDI resource characterization protocols and other MDI protocols that do not directly involve the characterization of specific resources, such as MDI-QKD.

In this work, we propose a framework that addresses all the questions mentioned above. We rigorously define the concept of MDI resource characterization protocols by introducing a quantum game involving two agents, Alice and Eve, resembling the experimentalist and quantum device in a quantum experiment. We focus on resource theories where local operation and shared randomness (LOSR) operations are free. We prove that MDI characterization protocols always exist for these resources and provide two general workflows to construct MDI protocols, which we name expectation-based and optimization-based protocols. LOSR operations consist the most general type of operation for spacelike separated systems, therefore our framework covers a wide range of resource theories related to quantum correlations and quantum nonlocality \cite{EPRSteering,NonlocalTeleportation,BuscemiNonlocality}, as well as different input/output types \cite{LOSR1,LOSR2}. Using the proposed framework, we generalize, unify, and simplify many existing MDI protocols and propose new ones, including the MDI classification of multipartite entanglement equivalence classes, the MDI estimation of multipartite entanglement for arbitrary resource monotones, and the MDI quantification of channel quantumness through channel-state duality, etc. We also present MDI protocols for nonlinear entanglement witnessing, removing the requirement for quantum memories or post-selection, thereby making the protocols much more applicable for near-term experiments while eliminating loopholes introduced by imperfect measurement
devices. More importantly, we demonstrate that MDI-QKD is indeed an MDI resource characterization protocol. By extracting secret key from the protocol, we achieve an estimation of the lower bound of the quantumness of the communication channel involved. This result emphasizes the fact that quantum memories or repeaters are fundamental resources in QKD protocols, and MDI-QKD characterizes them without the need to trust measurement apparatuses.

Our framework bridges the gap between resource theories of quantum nonlocality and MDI quantum information processing. On one hand, our findings shed light on the secure practical characterization of these nonlocality resources. On the other hand, we offer a perspective from resource theories to the study of MDI quantum information processing protocols, providing instructions for constructing new protocols that are MDI.

The paper is organized as follows: In Section \ref{Prelim}, we introduce the notations and preliminaries for this work, including a brief overview of the tensor network representation of quantum theory and existing resource theories relevant to this work. In Section \ref{GeneralFramework}, we establish the general framework of resource characterization protocols. In Section \ref{StaticEntanglement}, we apply our framework to the resource theories of static quantum entanglement. In Section \ref{QuantumMemories}, we apply our framework to the resource theory of quantum memories, or non-entanglement-breaking quantum channels. We leave most of the proofs to the appendices.

\section{Notations and preliminaries}
\label{Prelim}

This section is dedicated to the notations and preliminaries relevant to this work. 

\subsection{Basic notations and definitions}

We summarize the notations in Table \ref{TableNotations}. 

\begin{enumerate}
    \item Physical systems are denoted by capital letters, which will appear as superscripts throughout the paper.
    \item The convex set of all POVMs on system $A$ with outputs from $\mathcal{X}$ is denoted as $\mathscr{M}(\mathcal{H}^A;\mathcal{X})$. Its convexity arises from the definition of mixing two different $\mathcal{X}$-valued POVMs $\Pi^A,\Xi^A\in\mathscr{M}(\mathcal{H}^A;\mathcal{X})$: $\lambda \Pi^A + (1-\lambda) \Xi^A := (\lambda \Pi_x^A+(1-\lambda)\Xi_x^A;x\in\mathcal{X})$. We can also define the tensor product of two POVMs: $\Pi^A\otimes\Xi^A:=(\Pi^A_x\otimes \Xi^B_y;x\in\mathcal{X},y\in\mathcal{Y})\in \mathscr{M}(\mathcal{H}^A\otimes \mathcal{H}^B;\mathcal{X}\times\mathcal{Y})$. A POVM is called trivial when all its elements are proportional to the identity operator. A trivial $\mathcal{X}$-POVM can be implemented by discarding the quantum state and output elements in $\mathcal{X}$ according to a classical random number generator uniformly.
    \item We define resource theories as quadruples $\mathcal{R}=(\mathcal{G},\mathcal{F},\mathcal{O},\mu)$, where $\mathcal{F},\mathcal{O},\mu$ are free objects, free operations and resource monotone, respectively. We specify the set $\mathcal{G}$, which comprises all quantum objects considered in the resource theory. For example, in static quantum resource theories, $\mathcal{G}=\mathcal{D}(\mathcal{H}^A)$, while in dynamical quantum resource theories, $\mathcal{G}=\mathcal{Q}(\mathcal{H}^A\to\mathcal{H}^B)$. We require that $\mathcal{O}\subset\mathcal{B}(\mathcal{G}\to\mathcal{G})$, which differs from the literature like \cite{QRTReview,liu2019resourcetheoriesquantumchannels}, where free operations are defined as mappings between arbitrary physical systems. No generality is lost in this case, as we can always take $\mathcal{G}$ to include all ancillary systems involved. We also require the existence of a resource monotone $\mu$ for the resource theory. There are various axiomatizations for resource monotones, but we impose only the minimal requirements on $\mu$, which is (1) Nonnegativity: $\mu(g)\ge 0,\forall g\in\mathcal{G}$ and $\mu(f)=0,\forall f\in\mathcal{F}$; (2) Monotonicity: $\mu(o(g))\le \mu(g),\forall g\in\mathcal{G},o\in\mathcal{O}$. When necessary, we will impose additional properties for resource monotone, which are (3) Continuity: $\lim_{n\to\infty}|\mu(g_n)-\mu(g)|=0$ for every sequence of objects $\{g_n\}$ that satisfy $\lim_{n\to\infty}d(g_n,g)=0$. Here $d$ is a distance measure on $\mathcal{G}$; (4) Convexity: $\mu(\sum_i p_i g_i)\le \sum_ip_i\mu(g_i)$. 
\end{enumerate}

\begin{table}[h]
\centering
\caption{List of notations and abbreviations used in the paper}
\begin{tabular}{c|l}
\hline
\textbf{Notation} & \textbf{Description} \\ 
\hline
$A,B$               & Physical systems \\
$\mathcal{H}^A$               & Hilbert space with respect to a physical system A \\ 
$\mathcal{B}(\mathcal{H}^A)$            & Set of linear bounded operators on $\mathcal{H}^A$ \\ 
$\langle X,Y\rangle_{HS}$      & The Hilbert-Schmidt inner product of $X,Y\in\mathcal{H}$ \\ 
$\langle \mathcal{W}\rangle_\rho$ & Expectation value of $\mathcal{W}$ on $\rho$\\
$\mathcal{D}(\mathcal{H}^A)$      & Set of density operators as a subset of $\mathcal{B}(\mathcal{H^A})$ \\ 
$\mathcal{X},\mathcal{Y}$      & Sample spaces \\ 
$x,y$     & Instances taken value from $\mathcal{X},\mathcal{Y}$ \\ 
$\rho,\sigma$     & Quantum states, denoted by lower case Greek letters\\ 
$\tilde{\rho}$   & Unnormalized quantum state \\ 
$I^A$        & Identity matrix in $\mathcal{B}(\mathcal{H}^A)$ \\ 
$\Phi_+^{AA'}$            & Maximally entangled state $\frac{1}{d}\sum_{jk}\ketbra{jj}{kk}$ on $\mathcal{D}(\mathcal{H}^A\otimes\mathcal{H}^{A'})$\\ 
$\Pi_x^A$   & POVM element acting on $A$, with outcome $x\in\mathcal{X}$ \\ 
$\Pi^A$ & A $\mathcal{X}$-valued POVM on $A$, $\Pi^A=\{\Pi^A_x;x\in\mathcal{X}\}$    \\ 
$\mathscr{M}(\mathcal{H}^A;\mathcal{X})$     & Convex hull of all $\mathcal{X}$-valued POVMs on $A$ \\ 
$\mathcal{B}(\mathcal{H}^A\to\mathcal{H}^B)$     & Set of bounded linear maps from $\mathcal{B}(\mathcal{H}^A)$ to $\mathcal{B}(\mathcal{H}^B)$\\ 
$\mathcal{Q}(\mathcal{H}^A\to\mathcal{H}^B)$         & Set of quantum channels as a subset of $\mathcal{B}(\mathcal{H}^A\to\mathcal{H}^B)$ \\ 
$\mathcal{E}^\dagger$ & Adjoint map of a linear map $\mathcal{E}$ \\
$id^A$ & Identity map in $\mathcal{B}(\mathcal{H}^A\to\mathcal{H}^B)$\\ 
$J_\Lambda$ & Normalized Choi state associated to a quantum channel $\Lambda$ \\
$\mathcal{R}$ & Quantum resource theory as a tuple $(\mathcal{G},\mathcal{F},\mathcal{O},\mu)$ \\
$\mathcal{G}$ & Set of all objects considered in $\mathcal{R}$\\ 
$\mathcal{F}$ & Set of free objects in $\mathcal{R}$\\ 
$\mathcal{O}$ & Set of free operations in $\mathcal{R}$ \\ 
$\mu$ & resource monotone for $\mathcal{R}$\\ 
MDI & Measurement-device-independent \\ 
LOSR & Local operation and shared randomness \\ 
SEP & Separable states \\
EB & Entanglement-breaking \\
\hline
\end{tabular}
\label{TableNotations}
\end{table}

In this work, we mainly focus on a type of quantum channel called local operation and shared randomness (LOSR) \cite{LOSR1}. It is originally defined as (we demonstrate with the bipartite case for simplicity) a quantum channel $\mathcal{E}\in\mathcal{Q}(\mathcal{H}^A\otimes \mathcal{H}^B\to \mathcal{H}^{A'}\otimes \mathcal{H}^{B'})$ that can be expressed as $\mathcal{E}=\sum_\mu\pi(\mu)\mathcal{A}_\mu\otimes\mathcal{B}_\mu$, where $\mathcal{A}_\mu\in\mathcal{Q}(\mathcal{H}^A\to\mathcal{H}^{A'})$ and $\mathcal{B}_\mu\in\mathcal{Q}(\mathcal{H}^B\to\mathcal{H}^{B'})$ for all $\mu$, and $\pi(\mu)$ is a probability distribution corresponding to the shared randomness. It can be shown that LOSR operations are a subset of local operation and classical communication (LOCC) \cite{LOSR3}, which allows local operations and unlimited rounds of classical communication between the parties. One can also define LOSR superchannels as quantum superchannels where the pre- and post- processing channels \cite{QuantumSupermaps} are LOSR channels.

\subsection{Tensor network representation of quantum theory}

We briefly introduce the tensor network representation of quantum theory, sometimes referred to as quantum tensor networks, as it provides a clear and visualized illustration of the proofs and protocols in this paper. Curious readers may refer to more comprehensive tutorials such as \cite{TensorNetworks1,TensorNetworks2}.

\begin{figure}[htbp!]
    \centering
    \includegraphics[scale=0.2]{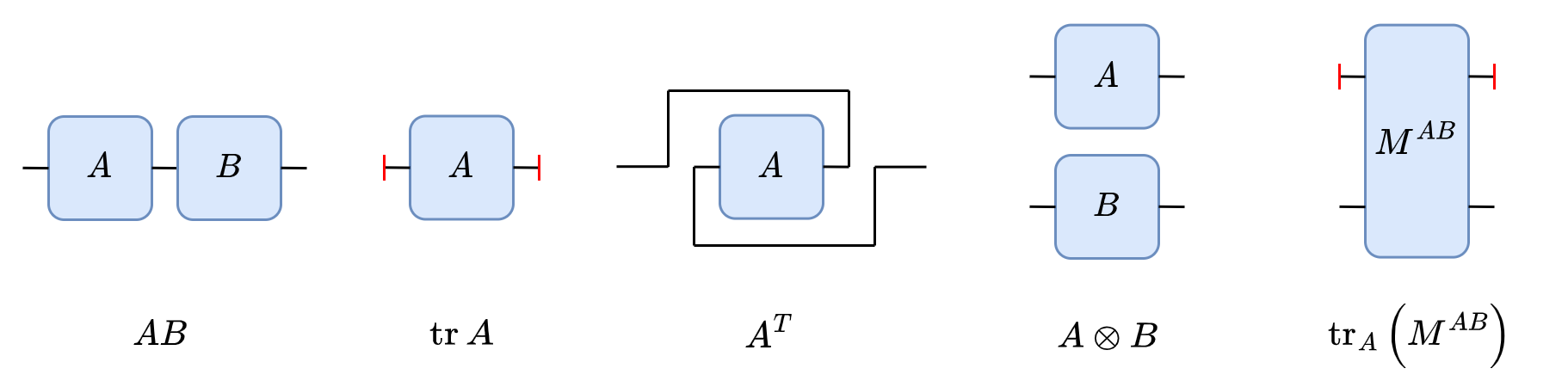}
    \caption{Tensor network representation of basic matrix operations.}
    \label{FigTensor1}
\end{figure}

In a tensor network representation, an object with uncontracted indices is depicted as a box with open legs. Row indices correspond to left-oriented legs, while column indices correspond to right-oriented legs. Connecting open legs stands for index contraction. Thus, matrix multiplication $AB$ is graphically represented by connecting the right legs of $A$ with the left legs of $B$. The tensor product of two matrices $A\otimes B$ does not contract any indices and is represented by placing the two boxes together. Taking the trace of a matrix $\tr A$ involves contracting the row and column indices of $A$, which can be represented by connecting the left and right legs of the box. Transposing a matrix $A^T$ corresponds to the action of switching the left and right indices. Similarly, for multipartite systems with more than one left and right oriented legs, we can define partial trace and partial transposition on a designated subsystem. The tensor network representation of basic matrix operations is summarized in Fig. \ref{FigTensor1}. 

We only contract indices that belong to the same Hilbert space, graphically corresponding to connecting open legs at the same horizontal level. Additionally, when taking trace on a given subsystem, we place a red vertical line on the left and right legs that need to be connected (which are required to be on the same horizontal level).

\begin{figure}[htbp!]
    \centering
    \includegraphics[scale=0.2]{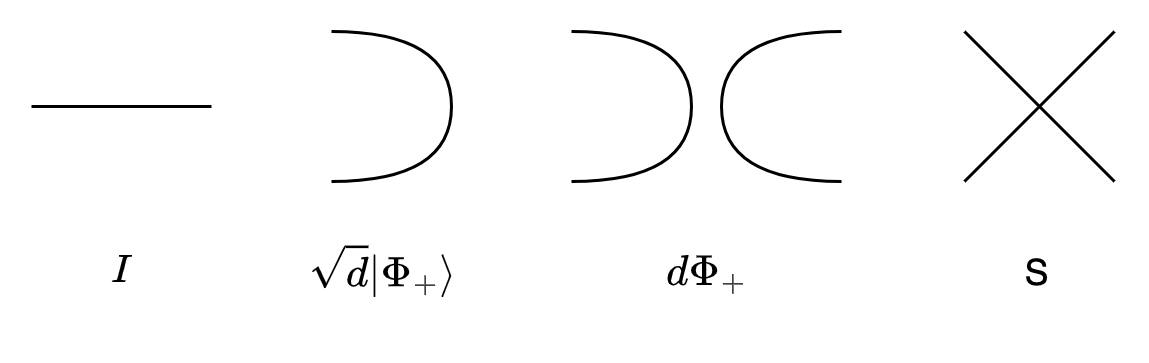}
    \caption{Tensor network representation of some states and operators.}
    \label{FigTensor2}
\end{figure}

We can also use tensor networks to represent states and operators beyond boxes. For example, the identity is represented as a simple straight line, while the unnormalized maximally entangled state $\sqrt{d}\ket{\Phi_+}$ is depicted as a semicircle. Similarly, the density matrix form of this state $d\Phi_+$ is composed of one semicircle pointing to the left and another pointing to the right. A similar operator is the swap operator, which interchanges the indices of the two systems it acts upon. Their graphical illustrations are shown in Fig. \ref{FigTensor2}.

\begin{figure}[htbp!]
    \centering
    \includegraphics[scale=0.2]{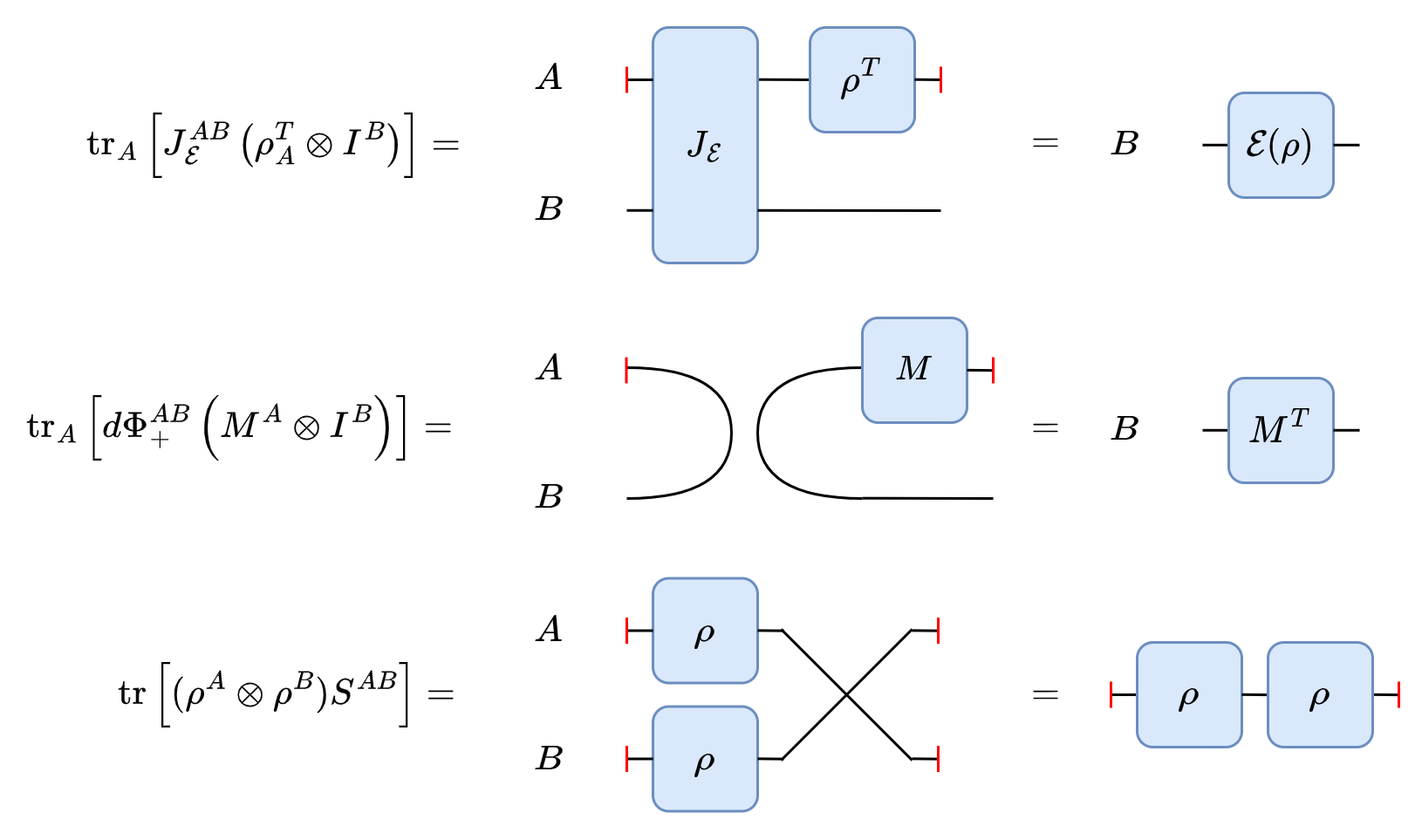}
    \caption{Tensor network representation of common subroutines.}
    \label{FigTensor3}
\end{figure}

Furthermore, we will label each horizontal level by the corresponding subsystem. We demonstrate this way of depiction by the examples shown in Fig. \ref{FigTensor3}. These examples are the commonly used subroutines in quantum information science and this work as well. 

\subsection{Examples of resource theories}

\subsubsection{Quantum entanglement}

We briefly introduce the resource theory of quantum entanglement. We begin with bipartite quantum systems, where the Hilbert space is $\mathcal{H}=\mathcal{H}^A\otimes\mathcal{H}^B$. The set of free states is called separable states and is defined as

\begin{equation}
    \mathcal{F}_{\text{SEP}}=\left\{\rho\in\mathcal{H}^{A}\otimes \mathcal{H}^{B}\left|\ \rho=\sum_\mu \pi(\mu)\rho^{A}_\mu\otimes\rho^{B}_\mu,\ \sum_\mu\pi(\mu)=1,\ \forall \mu,\ \pi(\mu)\ge0,\ \rho^{A}_\mu\in\mathcal{H}^{A},\ \rho^{B}_\mu\in\mathcal{H}^B\right.\right\}.
\end{equation}

Multipartite systems allow for richer entanglement structures to exist. If a quantum state can be written as a convex combination of states that are biseparable under different bipartitions, it is called a biseparable state or 2-separable state. More generally, the set of $k$-separable states is defined as 
\begin{equation}
    \mathcal{F}_{k}=\operatorname{conv}\left\{\rho\left|\right.\rho\in\mathcal{F}_{\left\{\mathfrak{A}_1\right\}\left\{\mathfrak{A}_2\right\}\cdots\left\{\mathfrak{A}_k\right\}}\right\},
    \label{kSEPstates}
\end{equation}
where $\operatorname{conv}$ stands for the convex hull of a given set, and
\begin{equation}
\begin{aligned}
\mathcal{F}_{\left\{\mathfrak{A}_1\right\}\left\{\mathfrak{A}_2\right\}\cdots\left\{\mathfrak{A}_k\right\}}=\left\{\rho\in\bigotimes_{j=1}^k\mathcal{H}^{\mathfrak{A}_j}\left|\ \rho=\sum_\mu\pi(\mu)\bigotimes_{j=1}^k\rho_\mu^{\mathfrak{A}_j},\sum_\mu\pi(\mu)=1,\ \forall\mu, \pi(\mu)\ge 0,\ \forall j, \ \rho_\mu^{\mathfrak{A}_j}\in\mathcal{H}^{\mathfrak{A}_j}\right.\right\},
\end{aligned}
\end{equation}
where the $k$-partition $\left\{\mathfrak{A}_1\right\}\left\{\mathfrak{A}_2\right\}\cdots\left\{\mathfrak{A}_k\right\}$ is an arbitrary partition of $\mathcal{H}$ involving $k$ parties. 

Many notions of multipartite entanglement can be described by the hierarchy of $\mathcal{F}_k$. For example, the set of fully separable states is given by $\mathcal{F}_{n}$. The set of quantum states with genuine multipartite entanglement is represented by $\mathcal{D}(\mathcal{H})\backslash\mathcal{F}_2$. The entanglement depth \cite{EntanglementDepth} of a multipartite quantum state is defined as the maximum number of (most finely grained) parties contained among all the decomposed reduced density matrices in the form of Eq.~\eqref{kSEPstates}.

The free operations in the resource theory of entanglement are LOCC operations, where the partites are allowed to perform local operations and classical communication.

Allowing unbounded classical communication enables the distribution of global randomness, demonstrating that LOSR operations are also free. The rigorous statement is given by the following proposition, which shows that the multipartite entanglement structure is preserved under LOSR transformations, with the proof in Appendix \ref{pfLOSRentanglement}:
\begin{proposition}
    Let $\mathcal{E}$ be an LOSR transformation acting on the Hilbert space $\mathcal{H}=\bigotimes_{j=1}^n \mathcal{H}^{A_j}$, with local operations being quantum channels acting on $A_j$. Then, for all $k=2,3,\cdots, n$, we have:
    \begin{equation}
        \mathcal{E}(\mathcal{F}_k):=\{\mathcal{E}(\rho)\left|\rho\in\mathcal{F}_k\right.\}\subseteq \mathcal{F}_k.
    \end{equation}
\label{LOSRentanglement}
\end{proposition}

Proposition \ref{LOSRentanglement} shows that in the context of resource theory, if we set the set of free objects to be $\mathcal{F}_k$, for any given $k$, then any LOSR operation will be resource non generating, making it suitable to be a free operation. 
\subsubsection{Quantum memories}

Now we focus on another quantum resource, which arises from the study of quantum memories \cite{QRTMemories,YuanMemories}. Briefly speaking, a quantum memory preserves the quantum information of the input state to the future. Such channels maintain the distinguishability of states. Classical memories cannot perform such operations, because by storing classical information from intermediate measurements, the entire channel has to be a measurement-and-prepare form:
\begin{equation}
\label{PMChannels}
\mathcal{N}^{A\to B}(\rho^A)=\sum_i \tr\left[\Pi^A_i\rho^A\right]\sigma_i^B,
\end{equation}
where $\Pi^A_i$ are POVM elements acting on system $A$ and $\sigma_i^B$ being fixed states. In the qubit case, if we want to preserve the distinguishability in the $Z$ basis, we are forced to perform measurements in the same basis, which will inevitably render $\ket{+}$ and $\ket{-}$ indistinguishable. Therefore, it is reasonable to designate channels of the form in  Eq.~\eqref{PMChannels} as free objects in the resource theory. 

It is well known that such channels form the set of EB channels \cite{EBChannels}. When a subsystem of a maximally entangled state $\Phi_+^{AA'}$ is sent through an EB channel $\mathcal{N}$, the output state, which is the Choi state of $\mathcal{N}$, is given by
\begin{equation}
    J_\mathcal{N}=(id^{A'}\otimes \mathcal{N}^{A\to B})(\Phi^{AA'}_+).
\end{equation}
This output state is separable. Therefore, resourceful objects are channels that possess an entangled Choi state.

Free operations can be simply defined as resource non-generating operations \cite{YuanMemories}, which are any physical transformations that map EB channels only to EB channels. 

There are many possible candidates of resource monotones. Here, we focus on a specific type of measure called the robustness \cite{GeneralResource} of quantum memories:
\begin{equation}
\label{Robustness}
\begin{aligned}
    \mu_R(\mathcal{N})&=\min_{\mathcal{M}\in\text{EB}}\left\{s\ge 0|\frac{\mathcal{N}+s\mathcal{M}}{1+s}\in\text{EB}\right\}\\
    &=\min_{J_\mathcal{M}\in\text{SEP}}\left\{s\ge 0|\frac{J_\mathcal{N}+sJ_\mathcal{M}}{1+s}\in\text{SEP}\right\},
\end{aligned}
\end{equation}
where $\text{EB}$ represents the set of all entanglement-breaking channels. The second equality follows from the Choi isomorphism. The robustness of quantum memories has many operational significance, including the tasks of memory synthesis and nonlocal games \cite{YuanMemories}. Eq.~\eqref{Robustness} shows that, in terms of robustness, one can quantify the quantumness of quantum memories by the amount of entanglement contained in their Choi states.

In summary, the resource theory we are primarily interested in can be formulated as the quadruple
\begin{equation}
    \mathcal{R}_{\text{MEM}}=\{\mathcal{Q}(\mathcal{H}^A\to\mathcal{H}^B),\text{EB},\text{RNG},\mu_R\},
    \label{QRTMEM}
\end{equation}
where RNG stands for resource non-generating operations.

\section{General framework}
\label{GeneralFramework}

In this section, we will introduce the general framework of MDI resource characterization protocols. Firstly, we will formulate a resource characterization protocol in general as a quantum game involving two agents, Alice and Eve, resembling the experimentalist and the quantum device in a quantum experiment. Secondly, we will define the notion of MDI resource characterization protocols by depicting Eve's manipulation power in the quantum game. Then we will show that every resource that is LOSR-free can be characterized by MDI protocols and explicitly provide workflows to construct them.

\subsection{Resource characterization protocol as a quantum game}

Upon obtaining an object that may be resourceful, its characterization task can be classified into two types: verification and quantification. 

A verification protocol is designed to determine whether a given object is resourceful. For example, in the case of entanglement witnesses, which is one of the most thoroughly studied verification protocols for quantum entanglement, the condition $\tr(\mathcal{W}\rho)< 0$ indicates entanglement with certainty, while $\tr(\mathcal{W}\rho)\ge 0$ leaves the problem uncertain. A quantification protocol follows a similar line of thought. Given a quantum device, we sometimes need to obtain or estimate a quantitative statement on the amount of resource it possesses. Here, we focus on the case where we aim to give a lower bound on the amount of resource.

To formalize this line of thought, we define the concept of resource characterization protocols.

\begin{definition}[Resource characterization protocol]
A $(\mathcal{R},c,s,m)$-resource characterization protocol for $g\in\mathcal{G}$ is a protocol designed to accept or reject the statement ``$\mu(g)> m$'', while satisfying the following properties:
\begin{enumerate}
    \item Completeness: If $\mu(g)>m$, the protocol rejects the statement with probability $c$.
    \item Soundness: If $\mu(g)\le m$, the protocol accepts the statement with probability $s$.
\end{enumerate}
\end{definition}

A resource characterization protocol can be viewed as a hypothesis testing scenario, where $m=0$ verifies the existence of the resource, while a protocol with $m>0$ estimates the lower bound of the amount of resource. The completeness and soundness parameters corresponds to type II and type I error rates. From now on, we will use the notion of resource characterization protocols instead of verification and quantification protocols.

Now we will place the hypothesis testing problem in the context of realistic devices. In practical experiments, the experimental device might suffer from imperfections, malfunctions, or, in the worst case, be controlled by a malicious adversary. By assigning different levels of trust to the device, these protocols can be classified into different types, such as trusted-device, device-independent and measurement-device-independent. In all cases except for the trusted-device scenario, the experimentalist is aware that the classical information generated by the device is no longer faithful, but he or she is still required to accept or reject the statement.

\begin{figure}[htbp!]
    \centering
    \includegraphics[scale=0.2]{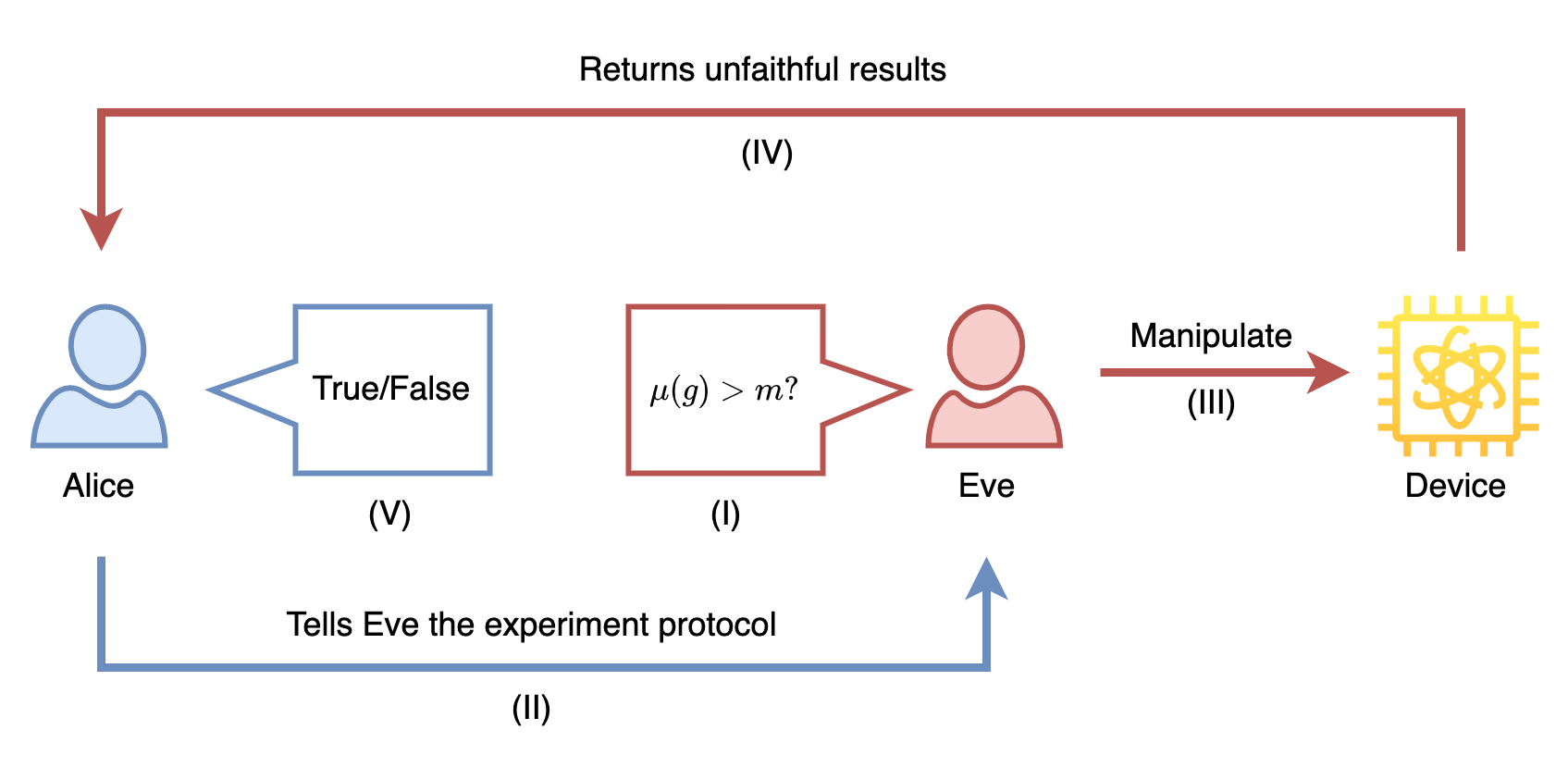}
    \caption{The schematic illustration of the game version of a resource characterization protocol. The five steps denoted by Roman numbers correspond to the procedures in Box \ref{BoxGame}. Blue/red arrows represent information that is trusted/untrusted by Alice.}
    \label{FigGame}
\end{figure}

We introduce a quantum game to describe the interplay between the experimentalist and the device. The game involves two players named Alice and Eve, along with an ideal quantum device that faithfully processes quantum information upon request. The procedures of the game are outlined in Box. \ref{BoxGame} and illustrated in Fig. \ref{FigGame}. 

\begin{mybox}[label={BoxGame}]{{Quantum game for resource characterization protocols}}
\begin{enumerate}[label=(\Roman*)]
	\item
	Eve draws $g$ from $\mathcal{G}$, which is an object known to Eve but unknown to Alice, and asks her whether $\mu(g)>m$. 
    \item 
    Alice tells Eve the experimental procedures that she intends to perform. An example of such a procedure is described in Section \ref{Conversion}, where Alice requests Eve to measure the expectation values of several observables with an explicit workflow.
    \item
    Eve carries out the workflow that Alice provides on the faithful device. However, she is allowed to manipulate the procedure once. Her manipulation power corresponds to the protocol's level of device dependency.
    \item 
    Eve tells Alice the result of the device as a response to her request, which is generally untrusted by Alice.
    \item
    Upon receiving the results, Alice performs classical post-processing on them and decides whether to accept or reject the statement. Eve wins the game if Alice accepts the statement, while Alice wins if she correctly discerns the validity of the statement.
    
\end{enumerate}
\end{mybox}

We emphasize that Eve's winning condition is to make Alice accept the statement rather than to make her falsely decide the statement's validity. Therefore, this is not a completely adversarial scenario. The main justification for this is that when Eve is completely adversarial, in most cases of interest, if $\mu(g)>m$, Eve can always lead Alice to reject the statement (see discussions in Section \ref{SectionOpt}), rendering our quantum game impractical. Another remark is that the role of Alice of Eve is similar to the verifier and prover in scenarios such as interactive proofs systems and complexity classes like QMA \cite{QuantumProofs}.

Before we end this subsection, we would like to remark that the completeness and soundness parameters are described by probabilities, which can occur in three possible ways: 
\begin{enumerate}
    \item The protocol itself is probabilistic.
    \item The object $g\in\mathcal{G}$ is unknown to Alice. Therefore, it can be modeled as being sampled from a prior distribution.
    \item Finite-size effects in real-world experiments prevent us from obtaining perfect expectation values of observables.
\end{enumerate}
In this work, we do not consider probabilistic protocols and neglect finite-size effects.

\iffalse
Again taking the example of entanglement witness, which corresponds to the verification of the resource theory of bipartite entanglement. If we assume that all the experiment apparatuses are trusted, by definition the soundness parameter $s=0$. The completeness parameter is less studied. Nevertheless, it is further proven that [] if we model the object $g\in\mathcal{D}(\mathcal{H}_d)$ to be sampled from a distribution of mixed stated generated by the partial trace of a Haar random pure state, the success probability $1-c$ for a protocol consisting of $O(1)$ single copy entanglement witnesses decays double exponentially with respect of the system size. These results show that entanglement detection protocols based on witnesses can indeed fit in our framework.
\fi

\subsection{MDI resource characterization}

In the quantum game for resource characterization, Eve's manipulation power over Alice's intended experimental workflow corresponds to the protocol's level of device dependency. For example, if the experimental device in a protocol is fully trusted, it satisfies the following definition of trusted-device resource characterization protocols:
\begin{definition}[Trusted-device resource characterization protocol]
    A ($\mathcal{R},c,s,m$)-resource characterization protocol for $g\in\mathcal{G}$ is a trusted device resource characterization protocol if, in its quantum game description, Eve has no manipulation power.
\end{definition}

On the contrary, if the protocol is fully device-independent, then in the corresponding quantum game, Eve is granted maximal manipulation power, with the only restrictions arising from fundamental physics, such as the no-signaling condition between spacelike separated parties.

This work aims to provide a general framework for MDI protocols. Informally speaking, Eve has  limited manipulation power over Alice's proposed workflow, namely at the measurement end. Therefore, we should expect Eve to be capable of arbitrarily manipulating the POVM measurements involved in the protocol as long as she obeys the no-signaling conditions.

Therefore, we arrive at the formal definition of MDI resource characterization protocols:
\begin{definition}[MDI resource characterization protocol]
\label{defMDI}
    A ($\mathcal{R},c,s,m$)-resource characterization protocol for $g\in\mathcal{G}$ is MDI if, during the corresponding quantum game, instead of performing the local POVM measurements suggested by Alice
        \begin{equation}
            \bigotimes_{j=1}^n\Pi^{A_j}\in\mathscr{M}(\bigotimes_{j=1}^{n}\mathcal{H}^{A_j};\mathcal{X}_1\times\mathcal{X}_2\times\cdots\times\mathcal{X}_n),
            \label{AlicePOVM}
        \end{equation}
        Eve is capable of implementing a mixture of new local POVM measurements
        \begin{equation}
            \sum_\mu \pi(\mu)\bigotimes_{j=1}^n\Xi_\mu^{A_j}\in\mathscr{M}(\bigotimes_{j=1}^{n}\mathcal{H}^{A_j};\mathcal{X}_1\times\mathcal{X}_2\times\cdots\times\mathcal{X}_n),
            \label{EvePOVM}
        \end{equation}
        where $\Pi^{A_j},\Xi_{\mu}^{A_j}\in\mathscr{M}\left(\mathcal{H}^{A_j};\mathcal{X}_j\right)$, and $\pi(\mu)$ is a probability distribution of pre-shared randomness.
\end{definition}

\begin{figure}[htbp!]
    \centering
    \includegraphics[scale=0.2]{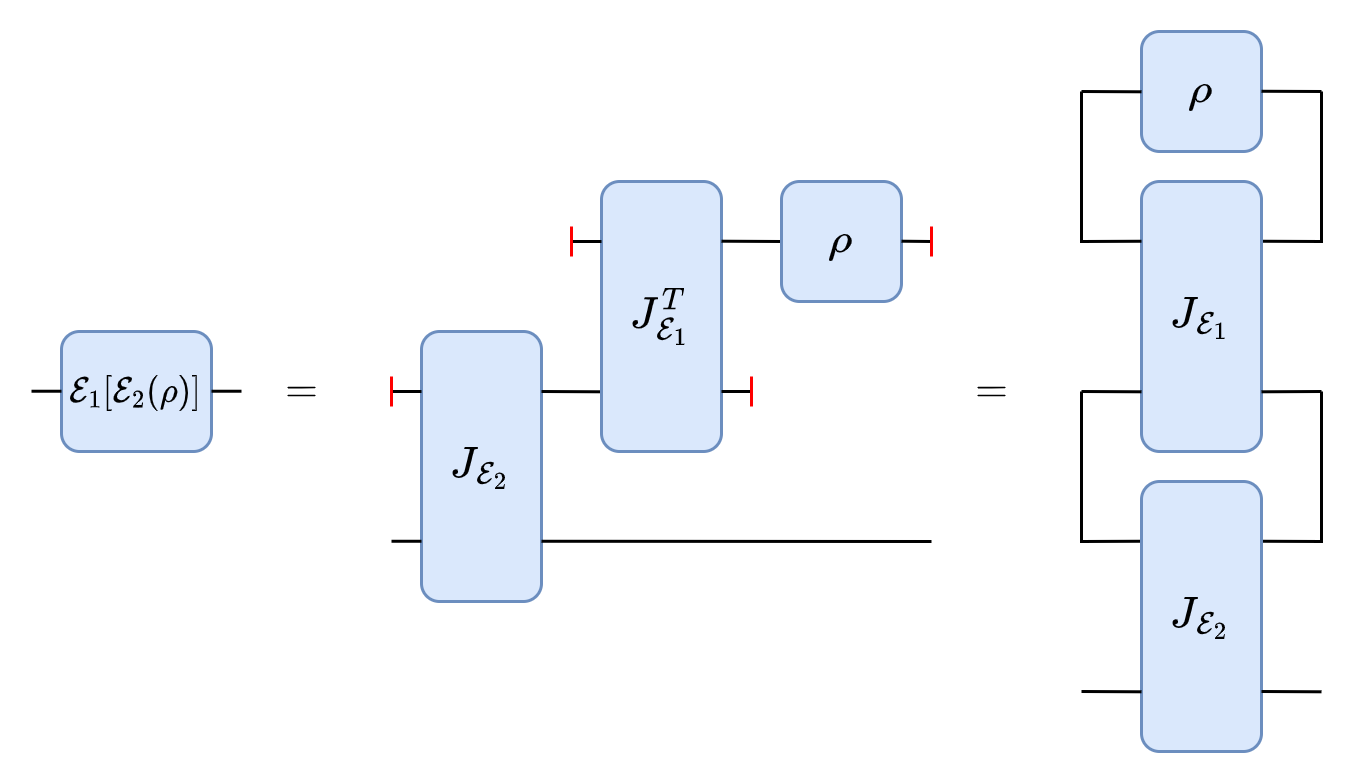}
    \caption{Example of transforming the action of quantum channels to static quantum states.}
    \label{FigChoi}
\end{figure}

Several remarks can be made regarding Definition \ref{defMDI}. 
\begin{itemize}
    \iffalse
    \item We require $\Pi^{A_j},\Xi_{\mu}^{A_j}$ are all $\mathcal{X}_j$-valued POVMs on the corresponding Hilbert space. This requirement can be justified due to the following reasoning: suppose Eve substitutes $\Pi^{A_j}$ into a $\mathcal{Y}_j$-valued POVM with $Y_j$ containing elements that do not exist in $\mathcal{X}_j$, Alice will receive outcomes that she had not expected and simply rejects the protocol. While if $\mathcal{Y}_j$ is strictly contained in $\mathcal{X}_j$, Alice equivalently encounters a $\mathcal{X}_j$-valued POVM with some outcomes occurs with probability zero. Therefore, no generality is lost from this requirement. 
    \fi
    \item The fact that $\Pi^{A_j}$ and $\Xi_{\mu}^{A_j}$ are POVMs of the same type is due to the premise that different subsystems (equivalently, the subscript $j$ is different) are spacelike separated. The spacelike separations forbid interactions between different subsystems, joint measurements between them, and classical communication between them for adaptive measurement strategies. We can always synchronize all spacelike separated measurements to occur at the same moment.
    \item 
    The spacelike separations between different parties also imply that we can synchronize all measurements, and there is intrinsically no need to consider time as a variable in the protocols. The reader might raise concerns about this claim because, in dynamical resource theories, temporal correlations arise from the use of quantum channels. Namely, two measurements can be performed at different times. Nevertheless, we can always exploit the channel-state duality and convert these dynamical processes into static quantum states, at the cost of introducing ancillary Hilbert spaces. A graphical demonstration of this conversion is presented in Fig. \ref{FigChoi}. Unless explicitly stated, we will adopt this perspective on dynamical resource theories. For example, the resource theory of quantum memories $\mathcal{R}_{\text{MEM}}$, defined in Eq. \eqref{QRTMEM},  will be mapped to the resource theory 
    \begin{equation}
        \{\mathcal{D}(\mathcal{H}^A\otimes\mathcal{H}^B),\mathcal{F}_{\text{SEP}},\text{RNG},\mu_R\}.
    \end{equation}
    In this context, resource non-generating operations are now separable operations \cite{SeparableOperation} that map separable states to separable states. 
    
    \item Notwithstanding, the requirement of spacelike separation can, in principle, be relaxed, allowing for classical communication among the subsystems. Details will be discussed in Appendix \ref{pfLOCC}, where we focus specifically on the resource theory of entanglement. However, in order to apply the framework to other possible resources, we will stay to this definition of measurement-device-independence in the main text.

\end{itemize}

\subsection{MDI protocols for LOSR-free resources}
\label{Conversion}

In this section, we prove that a general class of trusted-device characterization protocols for quantum resources, which do not increase under LOSR operations, can be transformed into MDI resource characterization protocols without losing completeness nor soundness. We also provide explicit protocols to achieve this transformation. Furthermore, we demonstrate that MDI characterization protocols exist for all LOSR-free resources. For clarity, we only focus on static resources. However, by applying the channel-state duality, this section is applicable to dynamical resources.

To begin with, we prove the local state decomposition lemma. To grasp the intuition behind this lemma, consider an $n$-qubit observable, which can be decomposed into an affine combination of Pauli strings. Pauli operators are not quantum states, since they are not positive semi-definite. However, if we utilize the substitution $\sigma_i=2\rho_i-I$, the observable can then be expressed as an affine combination of product states. For the most general case, we present the local state decomposition lemma, with the proof provided in Appendix \ref{pfStateDecomp}.

\begin{lemma}[Local state decomposition]
    Let $\mathcal{W}$ be a Hermitian observable acting on the $n$-partite Hilbert space $\mathcal{H}=\otimes_{j=1}^n \mathcal{H}^{A_j}$, where $\operatorname{dim} \mathcal{H}^{A_j}=d_{A_j}$. There exists a decomposition:
    \begin{equation}
        \mathcal{W} = \sum_{k_1=0}^{d_{A_1}^2-1}\sum_{k_2=0}^{d_{A_2}^2-1}\cdots \sum_{k_n=0}^{d_{A_n}^2-1}\beta_{k_1k_2\cdots k_n}\bigotimes_{j=1}^n \tau_{k_j}^{A_j},
        \label{DecomposedStateWithoutSuperscipt}
    \end{equation}
    where $\tau_{k_j}^{A_j}\in\mathcal{D}(\mathcal{H}^{A_j})$.
    \label{StateDecomp}
\end{lemma}

We remark that Lemma \ref{StateDecomp} provides only a proof of existence, and we do not claim that this decomposition is optimal. In fact, when $n=2$, there exists a decomposition based on operator Schmidt decomposition \cite{QRTMemories} that requires at most $\min\{d_{A_1},d_{A_2}\}^2+3$ decomposed terms. Unfortunately, since there is no Schmidt decomposition when $n\ge 3$, this approach cannot be directly generalized to multipartite systems. A search for optimal local state decomposition strategies is desirable for future research.

Now, we focus on a class of resource characterization protocols known as expectation-based protocols, which we define below:

\begin{definition}[Expectation-based resource characterization protocol]
    A resource characterization protocol is called expectation-based if it evaluates a function
    \begin{equation}
    C(\rho)=f\left[\tr(\mathcal{W}_1\rho^{\otimes N_1}),\tr(\mathcal{W}_2\rho^{\otimes N_2}),\cdots\tr(\mathcal{W}_M\rho^{\otimes N_M})\right],
    \label{OriginalProtocol}
    \end{equation}
    where $f$ is a multivariate function, and each observable $\mathcal{W}_k$ acts on $N_k$ identical copies of $\rho$.

    If $C(\rho)<0$, the protocol accepts the statement and rejects the statement vice versa.
\end{definition}

Most existing resource characterization protocols fall into this category, including entanglement witnessing \cite{Guhne2009,Witness}, quantum state learning models \cite{anshu_survey_2024}, and quantum tomography \cite{QubitTomo,QuditTomo}. 

\iffalse
We also prove the following lemma, which was initially conjectured in \cite{PracticalQuant}. The proof can be found in Appendix \ref{pfGlobalTransposition}.
\begin{lemma}
    In a static resource theory $\mathcal{R}$, if LOSR operations belong to the set of free transformations $\mathcal{O}$, the resource quantifier is invariant under global transposition, that is:
    \begin{equation}
        \mu(\rho)=\mu(\rho^T).
    \end{equation}
    \label{GlobalTransposition}
\end{lemma}
\fi

We are now ready to present our first theorem: 

\begin{theorem}[Existence of MDI protocols for expectation-based protocols]
If a static resource theory contains LOSR operations as free operations, and a transposition invariant resource monotone, that is $\mu(\rho)=\mu(\rho^T)$, then there exists a $(\mathcal{R},c',0,m)$-MDI resource characterization protocol for every expectation-based $(\mathcal{R},c,0,m)$-protocol, where $c'\le c$.
\label{MDIinGeneral}
\end{theorem}

\textit{Proof Sketch.} We provide a constructive proof by explicitly providing the workflow of the conversion in Box \ref{BoxConversion}:

\begin{mybox}[label={BoxConversion}]{{Constructing MDI protocols from expectation-based protocols}}
\begin{enumerate}[label=(\Roman*)]
	\item Take a Hermitian operator $\mathcal{W}$ from the $M$ operators. Suppose it acts non-trivially on a multipartite Hilbert space $\mathcal{H}=\otimes_{j=1}^n \mathcal{H}^{A_j}$ (which is a multiple copy of $\mathcal{G}$).  We apply the local state decomposition lemma and obtain the decomposed form given by Eq.~\eqref{DecomposedStateWithoutSuperscipt}.
    \item For each party $A$, assign an auxiliary party $A'$ with $\operatorname{dim}A'=\operatorname{dim}A=d_{A}$.
    
    \iffalse
    \item Denote one of the maximally entangled states on $\mathcal{H}_{A}\otimes \mathcal{H}_{A'}$ as $\ket{\Phi^+_{AA'}}=\frac{1}{\sqrt{d_{A}}}\sum_{k=1}^{d_A}\ket{kk}$, the generalized Bell basis on $\mathcal{H}_{A}\otimes\mathcal{H}_{A'}$ can be written as 
    \begin{equation}
    \{(I\otimes U_i)\ket{\Phi^+_{AA'}}\},\quad i=0,1\cdots d_A^2-1
    \end{equation}
    for a given set of local unitaries. 
    \fi
    
    \item For a complete unitary basis of $\mathcal{B}(\mathcal{H}^A)$, denoted by $\{U_i\}$, for each $i=0,1,\cdots ,d_{A}^2-1$, the operation $\tau_{k}^{A}\mapsto U_i\tau_{k}^{A}U_i^\dagger$ defines a non-singular linear transformation $T_i^A$ on a $d_{A}^2$-dimensional Hilbert space $\mathcal{B}(\mathcal{H}^A)$. Explicitly, we have the relation:
    \begin{equation}
    U_i\tau_{k}^{A}U_i^\dagger = \sum_{k'=0}^{d_{A}^2-1}\left(T^{A}_{i}\right)_{kk'}\tau_{k'}^{A}.
    \end{equation}
    Therefore, we can define an alternative decomposition of $\mathcal{W}$, parameterized by $i_1i_2\cdots i_n$:
    \begin{equation}
        \mathcal{W}= \sum_{k_1=0}^{d_{A_1}^2-1}\sum_{k_2=0}^{d_{A_2}^2-1}\cdots \sum_{k_n=0}^{d_{A_n}^2-1}\beta_{k_1k_2\cdots k_n}^{i_1i_2\cdots i_n}\bigotimes_{j=1}^n U_{i_j}\tau_{k_j}^{A_j}U_{i_j}^\dagger,
        \label{DecomposedStateWithSuperscript}
    \end{equation}
    where the coefficients are given by
    \begin{equation}
        \beta_{k_1k_2\cdots k_n}^{i_1i_2\cdots i_n}=\sum_{k'_1=0}^{d_{A_1}^2-1}\sum_{k'_2=0}^{d_{A_2}^2-1}\cdots \sum_{k'_n=0}^{d_{A_n}^2-1}\left[\prod_{j=1}^{n}\left(T^{-1}_{i_j}\right)^{A_j}_{k'_jk_j}\right]\beta_{k'_1k'_2\cdots k'_n}.
    \end{equation}
    \item Prepare quantum states $\omega^{A'_j}_{k_j}:=\left(\tau^{A_j}_{k_j}\right)^T$ for $k_j=0,1,\cdots ,d_A^2-1 $ on each auxiliary party $A'_j$. We assume that this process is carried out by trusted quantum sources that faithfully prepare these local states.
    \item Perform generalized Bell state measurements which need not to be trusted on the parties $A_jA'_j$ for all $j=1,2,\cdots ,n$. By a (faithfully performed) generalized Bell state measurement, we refer to a POVM measurement $\Pi^{AA'}=\{\Pi^{AA'}_i\}\in\mathscr{M}(\mathcal{H}^{AA'};\mathcal{X})$, where $\mathcal{X}=\{1,2,\cdots,d^2_A-1\}$, and each $\Pi^{AA'}_i$ is a rank-1 projector onto the generalized Bell state $\{(I\otimes U_i)\ket{\Phi_+^{AA'}}\}$. These measurements yield a set of probability distributions:
    \begin{equation}
        \left\{P(i_1,i_2,\cdots ,i_n|\omega^{A'_1}_{k_1},\omega^{A'_2}_{k_2},\cdots ,\omega^{A'_n}_{k_n})\right\}_{k_1,k_2,\cdots,k_n}^{i_1,i_2,\cdots,i_n}.
        \label{ProbabilityDistributions}
    \end{equation}
    \item Evaluate the MDI value:
    \begin{equation}
        \mathcal{I}(\rho) = \sum_{\substack{i_1=0 \\ k_1=0 }}^{d_{A_1}^2-1} \sum_{\substack{i_2=0 \\ k_2=0 }}^{d_{A_2}^2-1} \cdots \sum_{\substack{i_n=0 \\ k_n=0 }}^{d_{A_n}^2-1} \beta_{k_1k_2\cdots k_n}^{i_1i_2\cdots i_n} P(i_1,i_2,\cdots ,i_n|\omega^{A'_1}_{k_1},\omega^{A'_2}_{k_2},\cdots ,\omega^{A'_n}_{k_n}).
        \label{MDIvalue}
    \end{equation}
    The MDI value utilizes all possible POVM outputs in the index set $\mathcal{X}_1\times\mathcal{X}_2\times\cdots \times\mathcal{X}_n$.
    \item Repeating the procedure for all $M$ observables, we obtain a new function $C_{\text{MDI}}$:
    \begin{equation}
        C_{\text{MDI}}(\rho)=f\left[\Omega^{N_1}\mathcal{I}_1(\rho), \Omega^{N_2}\mathcal{I}_2(\rho),\cdots,\Omega^{N_m}\mathcal{I}_m(\rho)\right],
        \label{MDIProtocol}
    \end{equation}
    where $\Omega=\prod_{j=1}^n d_{A_j}$ is a constant.
    \item If $C_{\text{MDI}}(\rho)<0$, accept the statement. If $C_{\text{MDI}}(\rho)\ge0$, reject the statement.
    
\end{enumerate}
\end{mybox}

For the completeness parameter, we show that when $\mu(\rho)>m$, the strategy for Eve is to faithfully perform the generalized Bell state measurements that Alice intends to implement. Then, we have $C_{\text{MDI}}(\rho)=C(\rho)$, making the two protocols share the same completeness parameter.

For the soundness parameter, we can show that any manipulation by Eve will equivalently turn the MDI value into an expectation of $\mathcal{W}$ on a new quantum state. Specifically, we have: $\mathcal{I}_{\text{Eve}}(\rho)=\tr(\mathcal{W}^T\sigma_{\text{Eve}})=\tr(\mathcal{W}\sigma_{\text{Eve}}^T)$, where $\sigma_{Eve}$ is a quantum state that can be prepared from $\rho$ through LOSR operations. Therefore we have$\mu(\sigma^T_{\text{Eve}})=\mu(\sigma_{\text{Eve}})\le\mu(\rho)$. Since the soundness parameter is 0, $\sigma^T_{\text{Eve}}$ will also be rejected by the protocol with certainty.

\begin{figure}[htbp!]
    \centering
    \includegraphics[scale=0.125]{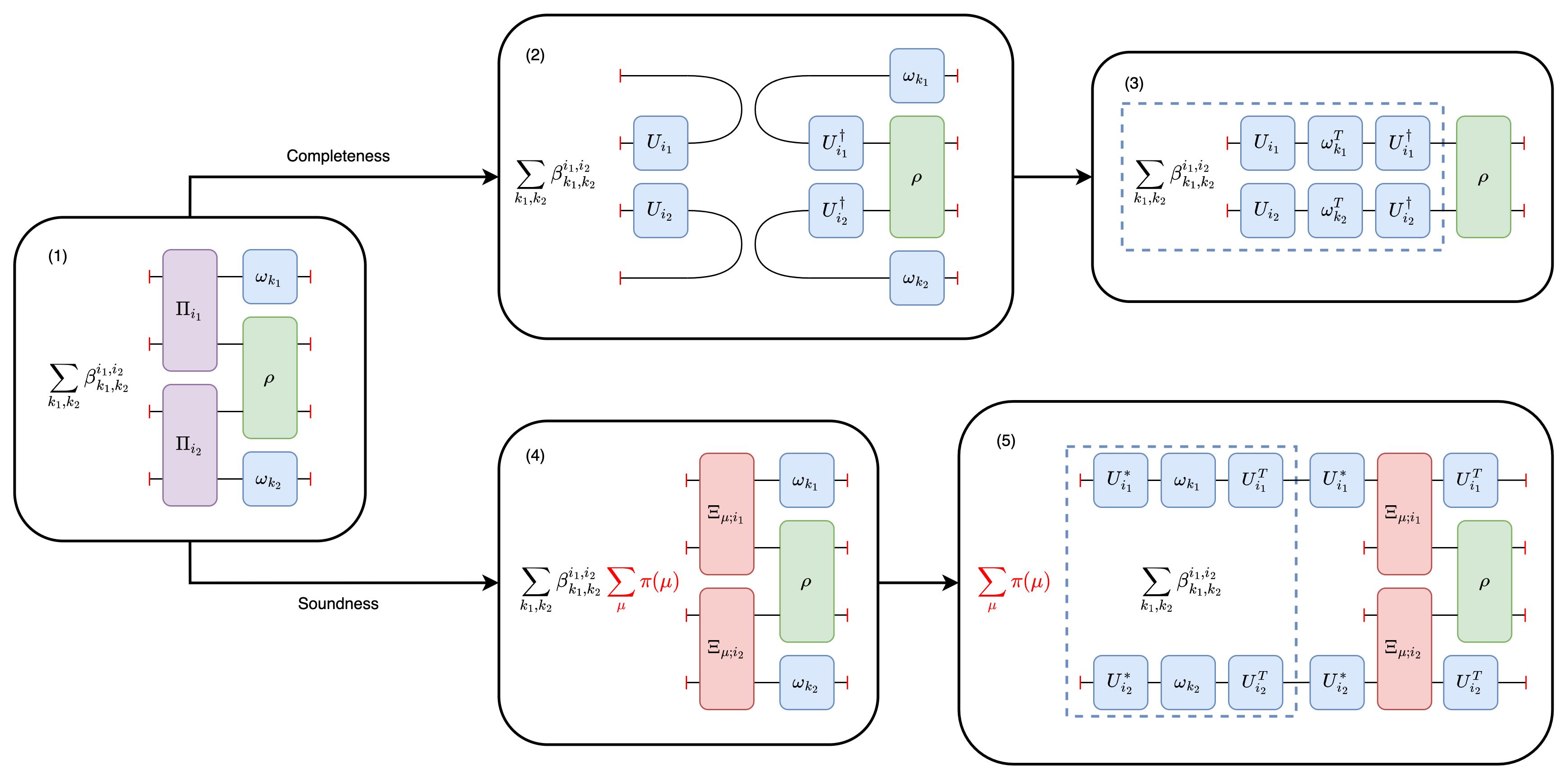}
    \caption{A graphical illustration of the proof. For clarity, we use a bipartite state as a case study. In the tensor network representation, blue denotes objects that are trusted or known by Alice, red indicates objects manipulated by Eve, and purple represents untrusted objects not yet manipulated by Eve. The state $\rho$ is marked in green. The tensor network in (1) represents the probability $P(i_1,i_2|\omega_{k_1},\omega_{k_2})$. When $\mu(\rho)>m$, a strategy for Eve is to follow the upper route to perform the measurement described by (2). It is evident that the expression in the dashed box in (3) is precisely $\mathcal{W}$, ensuring completeness at the same level. When $\mu(\rho)\le m$, any strategy by Eve falls into the lower route, as shown in (4). The dashed box in (5) corresponds to $\mathcal{W}^T$, while the remaining portion is an unnormalized quantum state that can be prepared from $\rho$ through LOSR operations.}
    \label{FigMDI}
\end{figure}

The details of the proof are presented in Appendix \ref{pfMDIinGeneral}, along with a schematic illustration of a case study shown in Fig. \ref{FigMDI}.

If we select a particular operator basis, a more explicit form can be constructed. In Appendix \ref{explicit}, we provide related calculations in detail and apply it to the case of two qutrits.

The following theorem shows that there exists an MDI resource characterization protocol for every LOSR-free resource, as long as the resource monotone is convex and continuous. A large class of general resource monotones satisfy this requirement, such as monotones based on distance between objects. 

\begin{theorem}[Existence of MDI protocols for every LOSR-free resource]
For any resource theory of the type $\mathcal{R}=(\mathcal{D}(\mathcal{H}),\mathcal{F},\mathcal{O},\mu)$ for some multipartite Hilbert space $\mathcal{H}$, where LOSR operations belong to $\mathcal{O}$ and $\mu$ is a continuous, convex and transposition invariant resource monotone, then there exists an $(\mathcal{R},c,0,m)$-MDI resource characterization protocol.
\label{EveryLOSR}
\end{theorem}

\textit{Proof.} We define a set
\begin{equation}
    \mathcal{F}_m:=\{\rho\in\mathcal{D}(\mathcal{H})|\ \mu_d(\rho)\le m\}
\end{equation}

The convexity of $\mathcal{F}_m$ comes from the convexity of $\mu$. Since $\mathcal{F}_m$ is the preimage of a closed set $[0,m]$ under a continuous function $\mu$, it is also a closed set in $\mathcal{D}(\mathcal{H})$. Notably, we do not require $\mathcal{F}$ to be a closed set, which may occur in infinite-dimensional Hilbert spaces \cite{kholevo2005notion}. 

Then, according to the Hahn-Banach theorem, there exist a hyperplane that separates a closed and convex set in a Hilbert space from a point that does not belong to this set. In other words, there exists a witness-like operator $\mathcal{W}_m$ such that $\tr(\rho \mathcal{W}_m)\ge 0$ for all $\rho\in\mathcal{F}_m$. Therefore, $\mathcal{W}_m$ corresponds to a $(\mathcal{R},c,0,m)$-resource characterization protocol. By applying Theorem \ref{MDIinGeneral}, we can construct an MDI characterization protocol based on $\mathcal{W}_m$. $\square$

\subsection{Optimization-based MDI protocols}
\label{SectionOpt}

Apart from the expectation-based resource characterization protocols described above, there is another approach to constructing MDI protocols based on optimization. Recall the quantum game illustrated in Fig. \ref{FigGame}. Alice knows that the feedback from Eve, which is some probability distributions, is untrusted because Eve has performed an unknown manipulation of the original protocol. So a possible strategy for Alice is to minimize $\mu(g)$ over all manipulations performed by Eve. The result is guaranteed to be MDI. 

\begin{figure}[htbp!]
    \centering
    \includegraphics[scale=0.2]{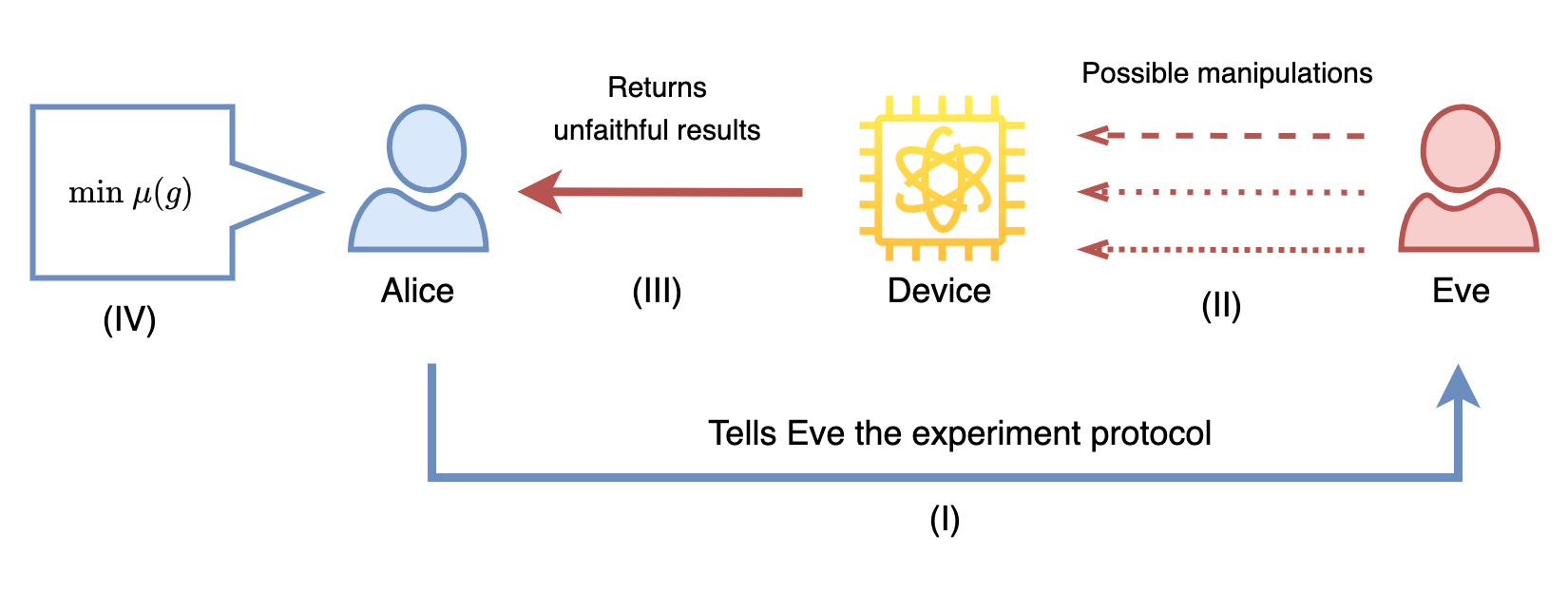}
    \caption{Schematic illustration of an optimization-based MDI resource characterization protocol. Similar to the original quantum game of general resource characterization protocols, it involves the interplay between Alice and Eve. (I) Alice specifies the experimental procedures she intends to perform. (II) Eve carries out the workflow, but the measurements are manipulated by her. The different arrows correspond to various manipulation strategies. (III) Eve returns the results of the device to Alice. (IV) Alice calculates the minimum of $\mu(g)$ over all $g$ that can produce the experimental results she receives.}
    \label{FigGame2}
\end{figure}

For the sake of clarity, we slightly modify the definition of resource characterization protocols when they are based on optimization. A schematic illustration of this is depicted in Fig. \ref{FigGame2}.

In a static resource characterization protocol, if we directly perform a single joint measurement on all parties of $\rho$, for instance, measuring the expectation of a multi-qubit observable by decomposing it into Pauli strings. Then the optimization problem can be formulated as follows:

\begin{equation}
\label{Opt}
    \begin{split}
        \min ~~& \mu(\rho) \\
        \text{over} ~~ & \Xi_i\ge 0, \forall i, ~~\sum_i \Xi_i=\bigotimes_{j=1}^nI^{A_j}, \\
        & \rho\ge 0,~~\tr(\rho)=1 \\
        s.t. ~~& \tr[\Xi_i\rho]=P(i),~~\forall i. 
    \end{split}
\end{equation}

However, the following proposition shows that the solution to this problem is always trivial:

\begin{proposition}
\label{OptTrivial}
    For all probability distributions described by $P(i)$, the outcome of the optimization problem in Eq.~\eqref{Opt} will always be 0.
\end{proposition}
\textit{Proof.} For all $\rho\in\mathcal{G}$, there is always a free state $\rho_f$ that can be converted to $\rho$ by a quantum channel $\mathcal{E}(\rho_f)=\rho$. Then we have
\begin{equation}
    \begin{aligned}
        P(i)&=\tr\left[\Xi_i \rho\right]\\
        &=\tr\left[\mathcal{E}^\dagger(\Xi_i)\rho_f\right].
    \end{aligned}
\end{equation}
Since the adjoint channel $\mathcal{E}^\dagger$ is completely positive and preserves the identity, the collection $\left\{\Xi_i\right\}_i$ is indeed another POVM that is applicable by Eve. Therefore, Eve can always create the same probability distribution from measuring a free object. $\square$

The operational meaning of this proposition is that for any resource characterization protocol performing a single joint measurement on all parties of $\rho$, there is always an attack on the measurement end that can counterfeit free objects to appear resourceful. Thus, optimization-based approaches such as \cite{QuantumData} cannot be useful when the measurement device is untrusted.

By the same experimental setting described in Box \ref{BoxConversion}, we can formulate a nontrivial version of the MDI optimization problem. Here we denote the resource monotone as $\mathcal{E}$: 

\begin{equation}
\label{MDIOpt}
    \begin{split}
        \min~~& \mathcal{E}(\rho) \\
        \text{over}~~&  \Xi^{A_jA'_j}_{\mu;i_j}\ge 0,~\forall j,i_j,\mu, ~~\sum_{i_j=0}^{d^2_{A_j}-1}\Xi^{A_jA'_j}_{\mu;i_j}=I^{A_j}\otimes I^{A'_j},\\
        & \rho^{A_1A_2\cdots A_n}\geq 0,~~\tr(\rho^{A_1A_2\cdots A_n})=1, \\
        & \pi(\mu)\ge 0,~\forall \mu, ~~\sum_\mu \pi(\mu)=1    \\
         s.t.~~& P(i_1,i_2,\cdots ,i_n|\omega^{A'_1}_{k_1},\omega^{A'_2}_{k_2},\cdots ,\omega^{A'_n}_{k_n})\\
         &=\tr\left[\left(\sum_\mu\pi(\mu)\bigotimes_{j=1}^n \Xi^{A_jA'_j}_{\mu;i_j}\right)\times\left(\rho^{A_1A_2\cdots A_n}\otimes \bigotimes_{j=1}^n\omega_{k_j}^{A'_j}\right)\right]
        ,~~\forall i,k.
    \end{split}
\end{equation}

 In the corresponding protocol, Alice inputs the ancillary states $\omega^{A'_1}_{k_1},\omega^{A'_2}_{k_2},\cdots ,\omega^{A'_n}_{k_n}$ and obtains measurement outcomes $i_1,i_2,\cdots ,i_n$. Upon receiving the probability distributions described by $P(i_1,i_2,\cdots ,i_n|\omega^{A'_1}_{k_1},\omega^{A'_2}_{k_2},\cdots ,\omega^{A'_n}_{k_n})$, she performs the optimization Eq.~\eqref{MDIOpt} and estimates the lower bound of $\mathcal(E)$. Unlike the optimization in Eq.~\eqref{Opt}, there exist probability distributions that are guaranteed to yield nontrivial results in Eq.~\eqref{MDIOpt}. 

 Interestingly, if the probability distribution is independent of the input ancillary states, for example,
 \begin{equation}
 P(i_1,i_2,\cdots ,i_n|\omega^{A'_1}_{k_1},\omega^{A'_2}_{k_2},\cdots ,\omega^{A'_n}_{k_n})=\frac{1}{\prod_{j=1}^n d^2_{A_j}},
 \label{TrivialProb}
 \end{equation}
 then every quantum state can potentially create such correlations, as long as Eve performs a trivial POVM on each subsystem. Therefore, the program in Eq.~\eqref{MDIOpt} will yield 0 if the probability distribution obtained is described by Eq.~\eqref{TrivialProb}.

 Therefore, as we have remarked before, we cannot set Eve as a complete adversary because she has the ability to force Alice to reject the protocol regardless of the underlying quantum object (by simply performing trivial POVMs).

By applying our established framework to various types of resource theories, we can design a plethora of MDI resource characterization protocols while at the same time generalizing, unifying, and simplifying existing quantum information processing protocols that are MDI.

\section{MDI characterization of entanglement}
\label{StaticEntanglement}

In this section, we will apply our framework to the characterization of entanglement, which has been proven to be an LOSR-free resource in Proposition \ref{LOSRentanglement}. We will discuss verification and quantification protocols, respectively.

\iffalse
The MDI verification of entanglement refers to the scenario where Alice and Bob want to verify whether the quantum state they share is entangled or not with untrusted measurement apparatuses, but with perfect state preparations. It has the property that once the measurement results indicate the state is entangled, then it must be entangled regardless of how the measurements are implemented. In other words, unfaithful implementations of measurement devices cannot identify a separable state as an entangled one, which makes the conclusion of entanglement more reliable. In Sec. \MakeUppercase{\romannumeral 3} A, we show how this works. In Sec. \MakeUppercase{\romannumeral 3} B, we show given the measurement results, what is the minimum amount of entanglement of the state without any knowledge on implementation details of measurement devices, namely, the MDI quantification of entanglement. From the perspective of free operations in quantum resource theory, we show that there are other quantum resources which can also be verified in an MDI way (Sec. \MakeUppercase{\romannumeral 3} C1). We also make clear some connections between MDI protocol and other verification protocols, such as nonclassical teleportation and quantum steering (Sec. \MakeUppercase{\romannumeral 3} C2).
\fi

\subsection{MDI entanglement verification protocols}

Here, we will provide general or novel constructions of MDI entanglement witnesses, which, by definition, are $(\mathcal{R}_{\text{ENT}},c,0,0)$-MDI resource characterization protocols. The soundness parameter is 0 because all separable states yield a nonnegative outcome, which can be rejected by the protocol with certainty. The completeness parameter depends on the probability distribution from which the state is drawn.

To summarize our novel results, by applying our framework, we proved that the linear entanglement witness for detecting multipartite entanglement classes can be made MDI, thereby addressing an open problem in the literature. We constructed an MDI nonlinear entanglement witness for arbitrary linear entanglement witnesses. The new witness protocol exhibits a stronger completeness parameter while remaining MDI. Compared to previous works on this approach, our protocol utilizes all possible measurement outcomes and does not require the input of maximally mixed states. We also provide a post-processing method for the measurement outcomes of single copy quantum states, which equivalently performs any multicopy entanglement witness at the MDI level. This result indicates that we can achieve the much stronger completeness of multicopy witnesses, such as universal entanglement witnesses for $2\times 2$ quantum systems, without relying on the use of quantum memories or higher-dimensional quantum systems that are experimentally expensive.

\subsubsection{Linear entanglement witness}

Linear entanglement witnesses are defined as expectation-based $(\mathcal{R}_{\text{ENT}},c,0,0)$-resource characterization protocols, with the multivariate function $C$ given by
\begin{equation}
    C(\rho)=\tr(\mathcal{W}\rho)
\end{equation}
for a given observable $\mathcal{W}$. 

Since the set of separable states is invariant under global transposition, by applying Theorem \ref{MDIinGeneral}, we can define an MDI version of any linear entanglement witness, which is a $(\mathcal{R}_{\text{ENT}},c',0,0)$-MDI resource characterization protocol with $c'\le c$, and the MDI value being

\begin{equation}
\label{MDIEW}
    C_{\text{MDI}}(\rho)=\Omega \mathcal{I}(\rho),
\end{equation}
where $\Omega$ and $\mathcal{I}(\cdot)$ are defined in Section \ref{Conversion}. In the literature, the protocol described by Eq.~\eqref{MDIEW} is referred to as a Measurement-Device-Independent-Entanglement-Witness (MDI-EW) \cite{Branciard2013}. Notice that by applying Theorem \ref{MDIinGeneral}, we automatically obtain a generalized version of MDI-EW that expands on the original protocol \cite{Branciard2013} in three aspects:
\begin{enumerate}
    \item It is applicable regardless of the dimensions of the subsystems. In other words, the dimensions of the parties can be arbitrary.
    \item It utilizes all outcomes of the (generalized) Bell measurements. The original protocol only relies on a specific outcome, which becomes inefficient when detecting high-dimensional or multipartite entanglement, as pointed out in \cite{Zhao2016}.
\end{enumerate}

Additionally, it can be shown that more entanglement properties can be detected by an MDI scheme. It is known that, apart from the entanglement structure denoted by entanglement depth, multipartite entanglement can be classified into different classes under stochastic local operation and classical communication (SLOCC). Conventional entanglement witnesses can be used to detect different entanglement classes, as shown in \cite{3QubitEnt}. It is stated as an open question in \cite{Zhao2016} to design MDI-EW from conventional witnesses for entanglement classification. Here, we answer this question affirmatively by simply noticing the fact that
\begin{equation}
    \text{LOSR}\subset\text{LOCC}\subset\text{SLOCC}.
\end{equation}

\begin{figure}[!htbp]
	\includegraphics[scale=0.15]{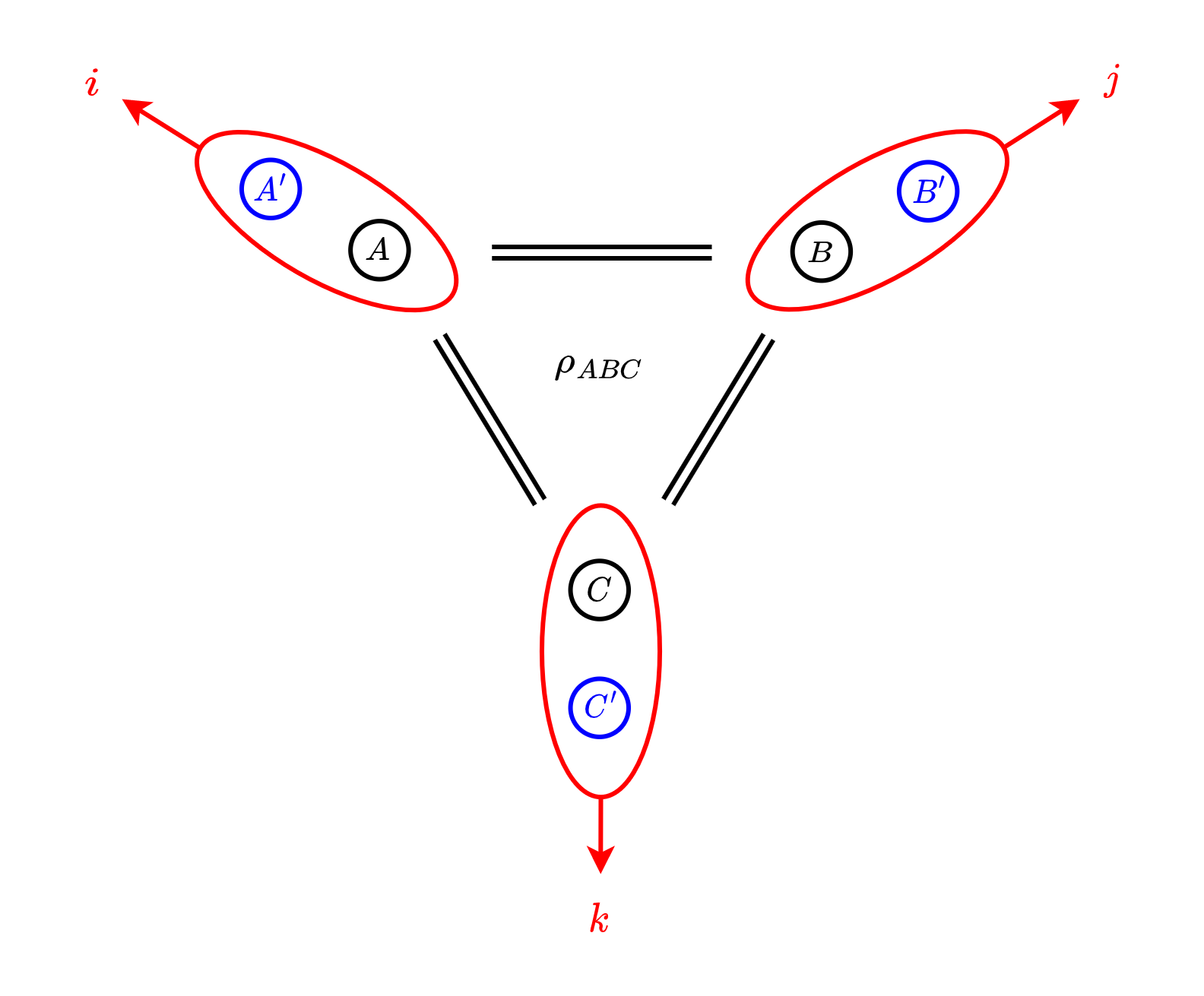}
    \caption{The setting for the MDI characterization of tripartite entanglement. Here, the black lines represent the quantum state $\rho^{ABC}$, the blue circles denote the ancillary input states, and the red circles indicate the joint measurements performed in the experiments. During an experiment, the experimenter obtains the probabilities of outcomes $i,j,k$ with respect to the quantum inputs, which are then used to characterize the tripartite entanglement.}
    \label{FigStaticEnt}
\end{figure}

Therefore, the SLOCC classes of entanglement are indeed unaffected by LOSR operations, indicating that Theorem \ref{MDIinGeneral} remains applicable. The experimental scheme of such protocols is depicted in Fig. \ref{FigStaticEnt} for the case of tripartite entanglement.

\subsubsection{Nonlinear entanglement witness}

We refer to the term nonlinear entanglement witness \cite{NonlinearWitness} as the procedure of finding a nonlinear correction to a given linear entanglement witness. According to our definition, it corresponds to an expectation-based protocol for entanglement, with the multivariate function defined as follows:
\begin{equation}
    C(\rho)=\tr(\mathcal{W}\rho)-\chi(\rho),
\end{equation}
where $\chi(\rho)$ is a nonlinear function of $\rho$.

By nonlinear correction, we require the new function to be strictly stronger than the original entanglement witness. Specifically, we have the conditions: $\forall \rho\in \text{SEP},C(\rho)\ge0$, and there exists $\sigma\in\text{ENT}$ such that $\tr(\mathcal{W}\sigma)\ge 0$ and $C(\sigma)<0$. Therefore, if the original linear entanglement witness corresponds to a $(\mathcal{R}_{\text{ENT}},c,0,0)$-expectation-based resource characterization protocol, the nonlinear entanglement witness corresponds to a $(\mathcal{R}_{\text{ENT}},c',0,0)$-resource characterization protocol that has a completeness parameter strictly less than $c$. 

Every entanglement witness can be related to its corresponding positive but not completely positive map \cite{Witness} denoted by $\Lambda^A$:
\begin{equation}
    \mathcal{W}=(\Lambda^A\otimes id^B)^\dagger(\ketbra{\phi}{\phi}),
\end{equation}
where the dagger represents the adjoint of a linear map, and $\ket{\phi}$ is the eigenstate of the map $(\Lambda^A\otimes id^B)$ corresponding to a negative eigenvalue, which always exists for every positive but not completely positive map.

It can be shown that for all positive but not completely positive map $\Lambda^A$, there are at least two constructions of $C(\rho)$ that can be used to create a nonlinear improvement of its corresponding entanglement witness \cite{NonlinearWitness}:

\begin{enumerate}
    \item Let $X=\ketbra{\phi}{\psi}$, where $\ket{\psi}$ is an arbitrary state, and let $s(\psi)$ denote the square of the largest Schmidt coefficient of $\ket{\psi}$. Then
    \begin{equation}
    \label{NEW1}
        C(\rho)=\langle \mathcal{W} \rangle_\rho-\frac{1}{s(\psi)}\langle (\Lambda^A\otimes id^B)^\dagger\ketbra{\phi}{\psi}\rangle_\rho\langle\left[(\Lambda^A\otimes id^B)^\dagger\ketbra{\phi}{\psi}\right]^\dagger\rangle_\rho
    \end{equation}
    is a nonlinear improvement of $\mathcal{W}$.
    \item Let $\{\ket{\psi_k}\}$ be an orthogonal basis on the joint Hilbert space. Then
    \begin{equation}
    \label{NEW2}
        C(\rho)=\langle \mathcal{W} \rangle_\rho-\sum_k \langle (\Lambda^A\otimes id^B)^\dagger\ketbra{\phi}{\psi_k}\rangle_\rho\langle\left[(\Lambda^A\otimes id^B)^\dagger\ketbra{\phi}{\psi_k}\right]^\dagger\rangle_\rho
    \end{equation}
    is a nonlinear improvement of $\mathcal{W}$.
\end{enumerate}

Both constructions depend on the evaluation of expectation values on $\rho$. Therefore, the nonlinear improvement of $\mathcal{W}$ is a $(\mathcal{R}_{\text{ENT}},c',0,0)$-expectation-based resource characterization protocol, and Theorem \ref{MDIinGeneral} still holds. Below, we provide the construction of an MDI nonlinear entanglement witness.

We begin with the form described in Eq.~\eqref{NEW2}, while constructions based on Eq.~\eqref{NEW1} are analogous. Notice that the non-Hermitian operator $(\Lambda^A\otimes id^B)^\dagger\ketbra{\phi}{\psi_k}$ can be decomposed as follows:
    \begin{equation}
        (\Lambda^A\otimes id^B)^\dagger\ketbra{\phi}{\psi_k}=H_k+iJ_k,
    \end{equation}
    where $H_k$ and $J_k$ are both Hermitian operators. Therefore, Eq.~\eqref{NEW2} can be expressed as
    \begin{equation}
    C(\rho)=\langle \mathcal{W}\rangle_\rho-\sum_{k} \langle H_k\rangle_\rho^2+\langle J_k\rangle_\rho^2.
    \end{equation}
    Next, we utilize the decomposition:
    \begin{equation}
        \begin{gathered}
        \mathcal{W} = \sum_{st}\beta^{ij}_{st}(U_i\tau_s^TU_i^\dagger)\otimes (V_j\omega_t^TV_j^\dagger),\\
        H_k = \sum_{st}\alpha^{ij}_{k,st}(U_i\tau_s^TU_i^\dagger)\otimes (V_j\omega_t^TV_j^\dagger),\\
        J_k = \sum_{st}\gamma^{ij}_{k,st}(U_i\tau_s^TU_i^\dagger)\otimes (V_j\omega_t^TV_j^\dagger).\\
        \end{gathered}
    \end{equation}
    Here, the ancillary states are chosen to be the same set, resulting in different expansion coefficients, and $U,V$ correspond to the unitaries that generate the generalized Bell states described in Box \ref{Conversion}.
    
    Then the MDI function we are going to evaluate becomes 
    \begin{equation}
    \label{MDINEW}
        C_{\text{MDI}}(\rho)=d_Ad_B\sum_{ijst}\beta_{st}^{ij}P(i,j|\tau_s,\omega_t)-d_A^2d_B^2\sum_{k}\left[\left(\sum_{ijst}\alpha_{k,st}^{ij}P(i,j|\tau_s,\omega_t)\right)^2+\left(\sum_{ijst}\gamma^{ij}_{k,st}P(i,j|\tau_s,\omega_t)\right)^2\right].
    \end{equation}
    According to Theorem \ref{MDIinGeneral}, this is indeed an MDI resource characterization protocol with a completeness parameter $c''\le c'$.

    Compared to the construction of MDI nonlinear witness in \cite{MDINEW}, our protocol utilizes all possible measurement outcomes. This will significantly lower the sample cost when the number of parties scales up \cite{Zhao2016}. Additionally, we do not require the input of maximally mixed states, which requires extra resources of unbiased randomness for precise generation.

\subsubsection{Multicopy entanglement witness}
\label{MulticopyWitness}

The previous discussion of nonlinear witnesses does not exhaust all possible expectation-based entanglement verification protocols with a zero soundness parameter. Specifically, we only discussed protocols that only require the evaluation of expectation values of single copy observables. Many more entanglement verification protocols exist if one allows for measuring observables acting on multiple copies of the quantum state.

\begin{figure}[htbp!]
    \centering
    \includegraphics[scale=0.2]{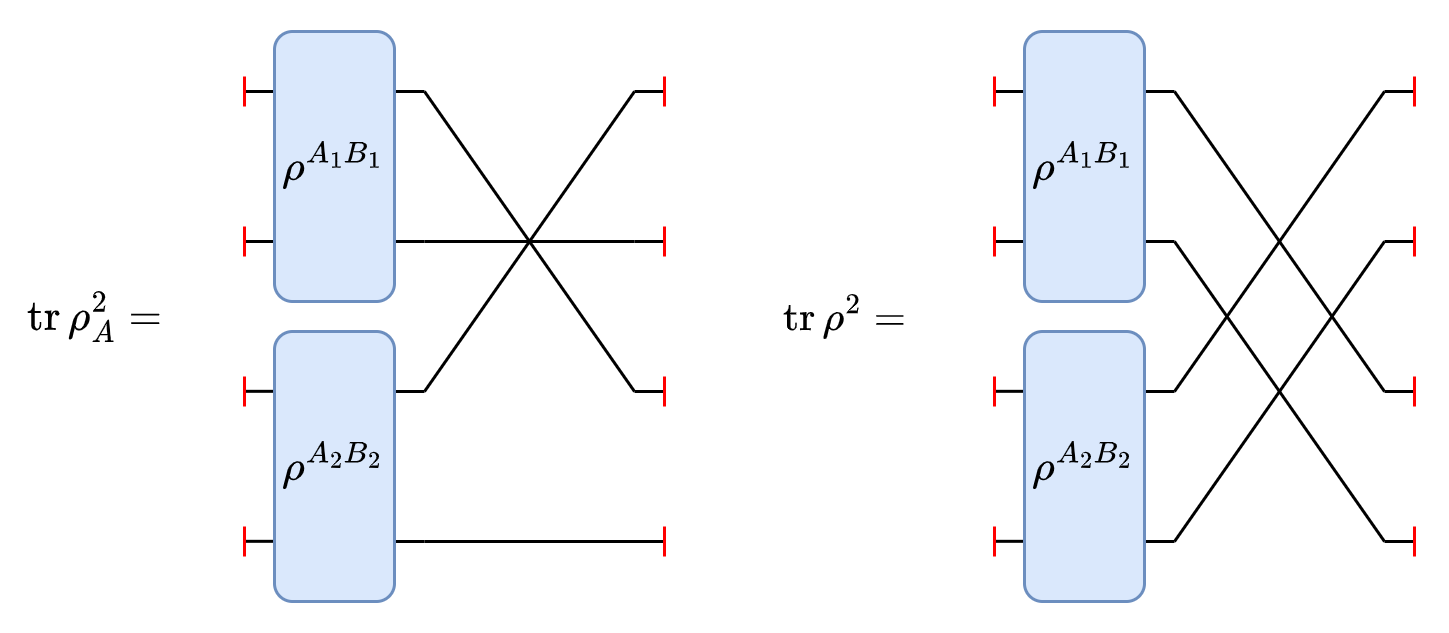}
    \caption{Illustration of the separability criterion by measuring expectation values of different swap operators}
    \label{FigSwap}
\end{figure}

As a simple example, according to the majorization criterion of separability \cite{MajorizationCriteria}, for any separable state $\rho\in\text{SEP}$, we have
\begin{equation}
    \tr(\rho^2_A)\ge\tr(\rho^2)
\end{equation}
for any subsystem $A$. Here $\rho_A$ refers to the reduced density matrix of subsystem $A$. This separability criterion can be viewed as an entanglement witness operator:
\begin{equation}
    \mathcal{W}^{(2)}=S^A\otimes I^B-S^{AB},
\end{equation}
where $S^A$ and $S^{AB}$ are swap operators depicted in Fig. \ref{FigSwap}.

Compared to single copy entanglement witnesses, multicopy witnesses have a significant advantage in terms of detection power. According to a recent result \cite{FundamentalLimitation}, for an $n$-qubit system coupled with an ancillary system, $\Omega(2^\wedge n/n)$ many single copy observables are required to detect the entanglement with $O(1)$ probability, but the same detection power can be achieved by using only one two-copy observable.

Also, for two-qubit quantum states, by utilizing the celebrated result that a state $\rho^{AB}$ is entangled if and only if $\det{\rho^{T_B}}<0$, where $\rho^{T_B}$ denotes the partial transposition of $\rho$ with respect to subsystem $B$, one can construct a universal entanglement witness acting on four copies of $\rho$ \cite{UniversalEW}. By "universal", we mean that the corresponding protocol becomes a $(\mathcal{R}_{\text{ENT}},0,0,0)$-resource characterization protocol, which cannot be achieved by witness protocols based on measuring expectations on one single copy observable.

\begin{figure}[!htbp]
	\includegraphics[scale=0.17]{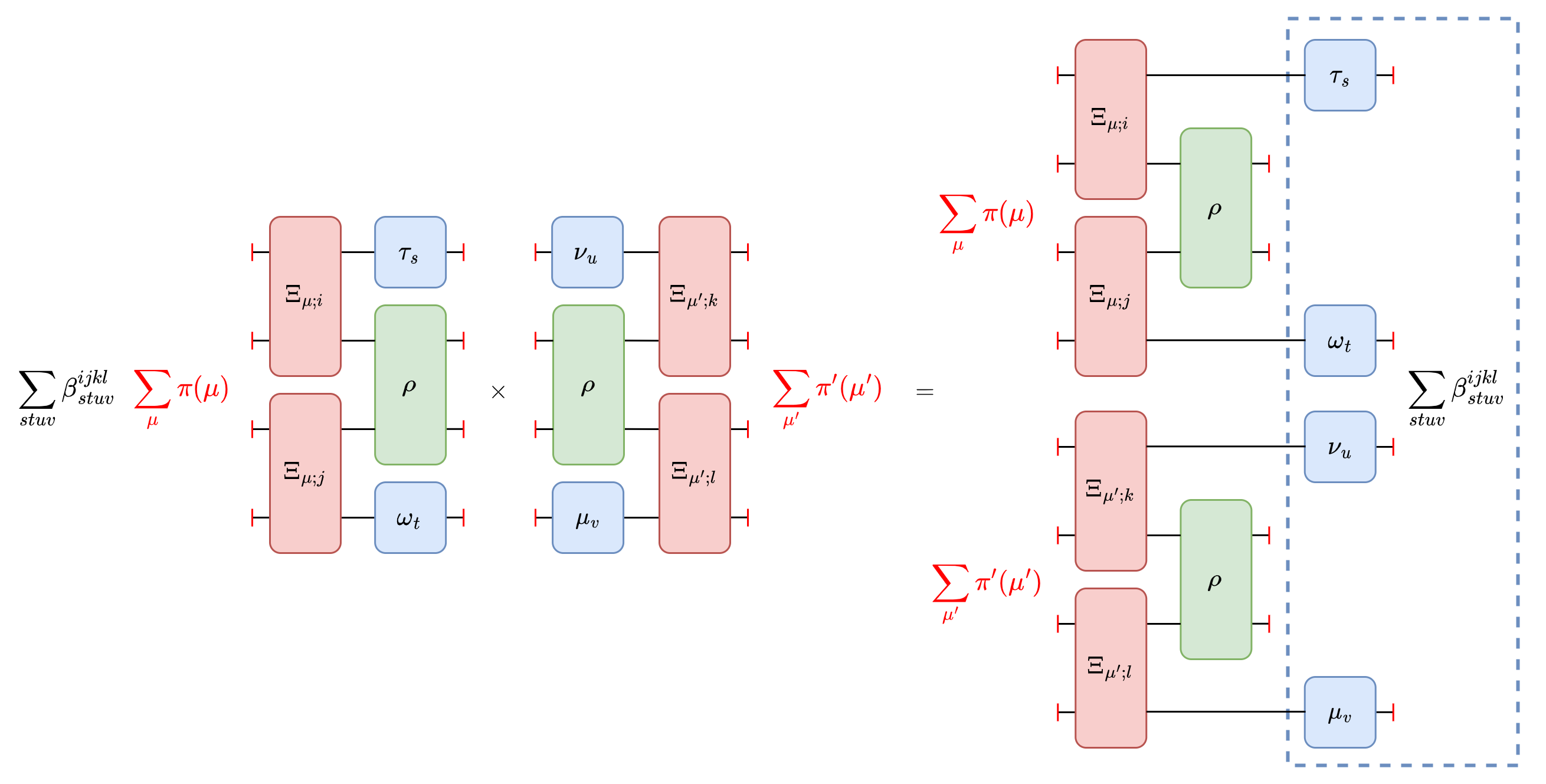}
    \caption{Illustration of an MDI multicopy entanglement witness. The convention of coloring is the same as Fig. \ref{FigMDI}. In the right hand side, the product of two different shared randomness $\pi(\mu)$ and $\pi'(\mu')$ is indeed another valid shared randomness by Eve. The dashed box corresponds to the decomposed operator $\mathcal{W}^{(2)}$.}
    \label{FigMulticopy}
\end{figure}

Given the importance of multicopy entanglement witnesses, we will explicitly show how they can be applied to our framework. As an example, we will use an arbitrary two-copy witness operator $\mathcal{W}^{(2)}$ that acts on the Hilbert space $\mathcal{H}^{A'}\otimes\mathcal{H}^{A}\otimes\mathcal{H}^{B'}\otimes\mathcal{H}^{B}$, where $A'$ and $B'$ refer to the other copies of the quantum state. By applying the local state decomposition lemma, we can obtain the decomposition of $\mathcal{W}^{(2)}$ in the form 
\begin{equation}
    \mathcal{W}^{(2)}=\sum_{stuv}\beta_{stuv}^{ijkl}(U_i\tau_s^TU^\dagger_i)\otimes (V_j\omega_t^TV_j^\dagger)\otimes (U_k\nu_u^TU_k^\dagger)\otimes(V_l\mu_v^T V_l^\dagger),
\end{equation}
where $U$ and $V$ are the local unitary transformations with respect to the generalized Bell states on $\mathcal{H}^A$ and $\mathcal{H}^B$. We can then continue the procedure described in Box \ref{BoxConversion}. 
An interesting observation is that, in this case, the probability distribution Eq.~\eqref{ProbabilityDistributions} can be further simplified as
\begin{equation}
    P(i,j,k,l|\tau_s,\omega_t,\nu_u,\mu_v)=P(i,j|\tau_s,\omega_t)P(k,l|\nu_u,\mu_v),
\end{equation}
if the measurements are faithfully performed. This suggests that we can remove the requirement of quantum memories to ensure that two copies of the state are present at the same time. This observation also holds if the measurements are manipulated by Eve. We show in Fig. \ref{FigMulticopy} that the individual attacks by Eve on two experiments are equivalent to a general attack on the protocol where two copies of the state is present in a single experiment, which is covered by Theorem \ref{MDIinGeneral}. Therefore, we have shown that every multicopy entanglement witness can be transformed into an MDI witness without the need for quantum memories, or equivalently, we only require a Hilbert space that supports a single copy of the unknown state, with the correlations of the multicopy witness encoded by classical post-processing. This result is in stark contrast to the MDI universal entanglement witness protocol proposed in \cite{MDIUEW}, which requires a much higher dimensional Hilbert space to encode multiple copies of the quantum state to be witnessed.

It is worth remarking that if we use a tomographically complete basis as the trusted quantum input, then this protocol can be viewed as the MDI version of quantum state tomography. It has the same sample complexity as standard state tomography but offers the further advantage of being robust against any measurement errors. Therefore, when we are only interested in obtaining information about the LOSR-free resources of an unknown state, it is desirable to use this protocol to enhance robustness and security.

\subsection{MDI entanglement quantification protocols}

In this section, we will focus on entanglement quantification protocols that aim to lower bound the amount of entanglement. Our novel results are twofold. Firstly, we analyze the problem of lower bounding the amount of entanglement from the expectation value of a single witness operator. In the trusted-device scenario, we review two existing methods and propose a new method based on witnessing the set of low entanglement states. We show that all of these methods can be transformed into MDI protocols. Secondly, we provide a general semidefinite program for MDI entanglement quantification that is capable of addressing multipartite cases and applies to any entanglement monotone.

\subsubsection{Lower bounds from entanglement witness}

We consider the problem of estimating the lower bound of entanglement from the expectation of a witness operator. That is, we are looking for a statement of the form
\begin{equation}
    \mathcal{E}(\rho)\ge f[\langle \mathcal{W}\rangle_\rho],
\end{equation}
where $\mathcal{E}$ is an entanglement monotone.

We first work in the trusted-device scenario. It is shown in \cite{QuantitativeWitness2} that for any entanglement witness $\mathcal{W}$, the trace distance based entanglement monotone
\begin{equation}
    \mathcal{E}_\text{tr}(\rho)=\min_{\sigma\in\text{SEP}} \frac{1}{2}\tr\|\rho-\sigma\|_1
\end{equation}
can be lower bounded by the function
\begin{equation}
    f_{\text{tr}}[\langle \mathcal{W}\rangle_\rho]=-\frac{\langle\mathcal{W}\rangle_\rho}{\lambda_+-\lambda_-},
\label{function1}
\end{equation}
where $\lambda_\pm$ are, respectively, the largest and smallest eigenvalues of $\mathcal{W}$.

Another approach is based on the Legendre transformation \cite{QuantitativeWitness1}. For any entanglement monotone $\mu$ that is convex and continuous, one can find an optimal bound described by the function
\begin{equation}
    f_{\text{opt}}[\langle\mathcal{W}\rangle_\rho]=\sup_\alpha\{\alpha\langle\mathcal{W}\rangle_\rho-\hat{\mu}(\alpha\mathcal{W})\},
\label{function2}
\end{equation}
where $\hat{\mu}$ is the Legendre transform for the witness operator $\mathcal{W}$, defined by
\begin{equation}
    \hat{\mu}(\mathcal{W})=\sup_\rho\{\tr(\mathcal{W}\rho)-\mu(\rho))\}.
\end{equation}
This bound given by $f_{\text{opt}}$ is optimal because there always exists a state that saturates the bound.

Finally, we have shown in our proof of Theorem \ref{EveryLOSR} that for arbitrary $m\ge0$, there exists a witness operator $\mathcal{W}_m$ that serves as a hyperplane, with every element in the set
\begin{equation}
    \mathcal{F}_m:=\{\rho\in\mathcal{D}(\mathcal{H})|\ \mathcal{E}(\rho)\le m\}
\end{equation}
lying on one side of the hyperplane. This result applies to every distance-based entanglement monotone $\mu_d$. Therefore, we can design another type of entanglement lower bound from the function
\begin{equation}
f_m[\langle\mathcal{W}_m\rangle_\rho]=\left\{
    \begin{aligned}
    &0 \qquad \langle\mathcal{W}_m\rangle_\rho\ge 0 \\
        &m \qquad \langle\mathcal{W}_m\rangle_\rho<0
    \end{aligned}\right..
\label{function3}
\end{equation}
It is straightforward to see from our Theorem \ref{MDIinGeneral} that all the witness involved in the functions above can be constructed to become MDI protocols, as long as the montone is invariant under transposition. In these cases, the variable $\langle\mathcal{W}\rangle_\rho$ being substituted with $\Omega\mathcal{I}(\rho)$. For any manipulation from Eve, we have $\Omega\mathcal{I}_{\text{Eve}}(\rho)=\langle\mathcal{W}\rangle_{\Lambda(\rho)}$ for some LOSR operation $\Lambda$. Equivalently, estimating the lower bound of $\mathcal{E}[\Lambda(\rho)]$ will also provide a lower bound for $\mathcal{E}(\rho)$.

\begin{figure}[!htbp]
	\includegraphics[scale=0.2]{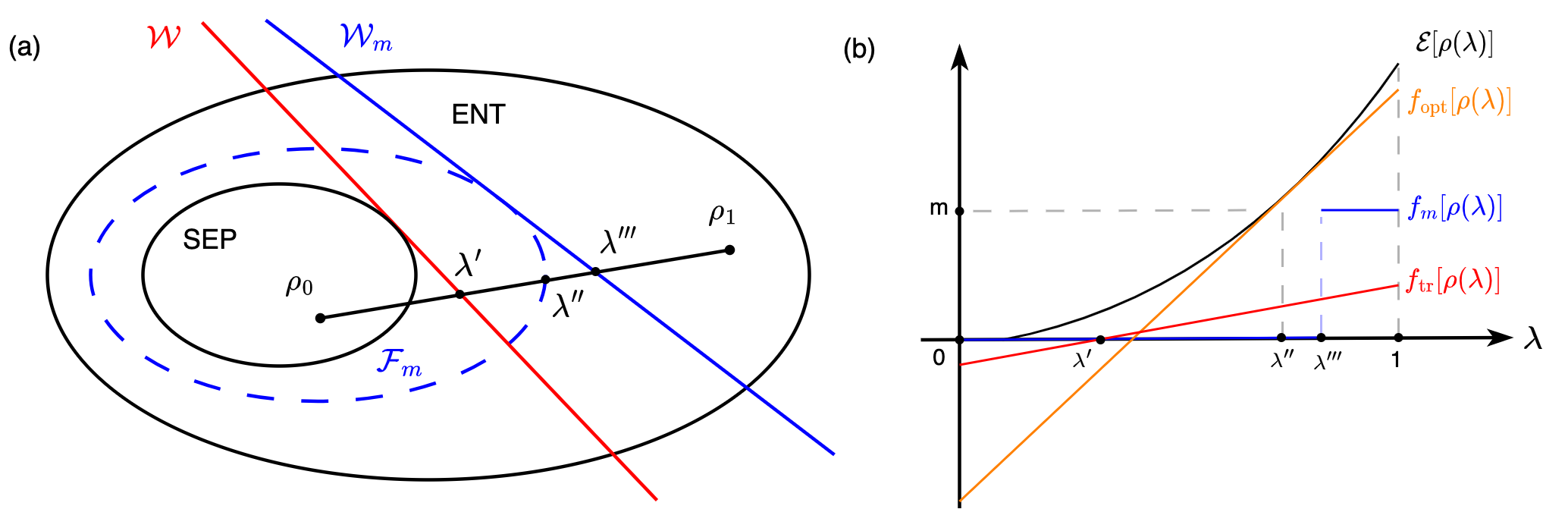}
    \caption{An illustration of methods that estimate the entanglement lower bound from a witness operator. For simplicity, the entanglement monotones involved are the measures  applicable to the current method, respectively. (a) The red line corresponds to a standard linear witness $\mathcal{W
    }$ with respect to separable states. The states with $\lambda>\lambda'$ can be detected by $\mathcal{W}$. The blue dashed curve depicts the boundary of $\mathcal{F}_m$, and the blue line corresponds to the witness operator $\mathcal{W}_m$ with respect to $\mathcal{F}_m$. The states with $\lambda>\lambda''$ have entanglement greater than $m$, and the states with $\lambda>\lambda'''$ can be detected by $\mathcal{W}_m$. (b) The plot of the three entanglement lower bounds. The function Eq.~\eqref{function1} is a linear function that is generally not tight. The function Eq.~\eqref{function2} is also a linear function but is optimal with respect to the obtained expectation value. The function Eq.~\eqref{function3} is a nonlinear function that guarantees entanglement of $m$ if it successfully witnesses a high entanglement state. Note that this plot is not quantitative.}
    \label{FigQuantWitness} 
\end{figure}

As an illustration, we pick a quantum state $\rho_0$ that is separable and another state $\rho_1$ that is entangled. Then the states lying on the segment connecting these two states can be described by a parameter $\lambda\in[0,1]$:
\begin{equation}
    \rho(\lambda)=(1-\lambda)\rho_0+\lambda\rho_1.
\end{equation}
We depict the geometrical representation and the three methods of entanglement lower bounds in Fig. \ref{FigQuantWitness}.

\subsubsection{Optimization-based entanglement quantification}

Another approach to entanglement quantification is based on semidefinite programming, which follows the notion of optimization-based resource characterization protocols presented in Section \ref{SectionOpt}.

We begin with the optimization problem directly following Eq.~\eqref{MDIOpt}:
\begin{equation}
    \begin{split}
        \min~~& \mathcal{E}(\rho) \\
        \text{over}~~&  \Xi^{A_jA'_j}_{\mu;i_j}\ge 0,~\forall j,i_j,\mu, ~~\sum_{i_j=0}^{d^2_{A_j}-1}\Xi^{A_jA'_j}_{\mu;i_j}=I^{A_j}\otimes I^{A'_j},\\
        & \rho^{A_1A_2\cdots A_n}\geq 0,~~\tr(\rho^{A_1A_2\cdots A_n})=1, \\
        & \pi(\mu)\ge 0,~\forall \mu, ~~\sum_\mu \pi(\mu)=1    \\
         s.t.~~& P(i_1,i_2,\cdots ,i_n|\omega^{A'_1}_{k_1},\omega^{A'_2}_{k_2},\cdots ,\omega^{A'_n}_{k_n})\\
         &=\tr\left[\left(\sum_\mu\pi(\mu)\bigotimes_{j=1}^n \Xi^{A_jA'_j}_{\mu;i_j}\right)\times\left(\rho^{A_1A_2\cdots A_n}\otimes \bigotimes_{j=1}^n\omega_{k_j}^{A'_j}\right)\right]
        ,~~\forall i,k,
    \end{split}
\end{equation}
where $\mathcal{E}$ is an entanglement monotone. The optimization provides a direct nontrivial MDI quantification of entanglement. For the sake of feasibility, we can further seek a lower bound of any entanglement monotone that can be more easily computed. We first extend the definition of entanglement monotones to unnormalized quantum states:
\begin{equation}
    \mathcal{E}(\tilde{\rho}):=\tr(\tilde{\rho})\mathcal{E}\left(\frac{\tilde{\rho}}{\tr(\tilde{\rho})}\right),\quad \mathcal{E}(0):= 0.
\end{equation}
Then we present the following lemma, with the proof in Appendix \ref{pfPOVMBound}:

\begin{lemma}
\label{POVMBound}
    For an arbitrary $n$-partite entanglement monotone $\mathcal{E}$ that is invariant under transposition, and for arbitrary $n$ POVMs $\Pi^{A_jA'_j}=\{\Pi^{A_jA'_j}_{i_j}\}\in\mathscr{M}(\mathcal{H}^{A_jA'_j};\mathcal{X}_j),~j=1,2,\cdots, n$, the following inequality holds:
    \begin{equation}
        \mathcal{E}(\rho)\ge \frac{1}{\Omega}\sum_{i_1\in\mathcal{X}_1}\sum_{i_2\in\mathcal{X}_2}\cdots\sum_{i_n\in\mathcal{X}_n}\mathcal{E}\left(\tilde{\rho}^{A'_1A'_2\cdots A'_n}_{i_1i_2\cdots i_n}\right),
    \end{equation}
    where $\mathcal{X}_j$ is an arbitrary index set with no more than $d_{A_j}^2$ elements, $\dim\mathcal{H}^{A_j}=\dim\mathcal{H}^{A'_j}=d_{A_j}$, $\Omega=\Pi_{j=1}^nd_{A_j}$, and
    \begin{equation}
        \tilde{\rho}^{A'_1 A'_2\cdots A'_n}_{i_1i_2\cdots i_n}=\tr_{A_1A_2\cdots A_n}\left[\left(\bigotimes_{j=1}^n \Pi^{A_jA'_j}_{i_j}\right)\times \left(\bigotimes_{j=1}^n I^{A'_j}\otimes\rho^{A_1A_2\cdots A_n}\right)\right]
    \end{equation}
    is an unnormalized quantum state.
\end{lemma}

With this relaxation, we provide an alternative program that belongs to an SDP, which also provides a lower bound of $\mathcal{E}(\rho)$:
\begin{equation}
    \begin{split}
        \min~~& \frac{1}{\Omega}\sum_{i_1\in\mathcal{X}_1}\sum_{i_2\in\mathcal{X}_2}\cdots\sum_{i_n\in\mathcal{X}_n}\mathcal{E}\left(\tilde{\rho}^{A'_1A'_2\cdots A'_n}_{i_1i_2\cdots i_n}\right) \\
        \text{over}~~&  \tilde{\rho}^{A'_1A'_2\cdots A'_n}_{i_1i_2\cdots i_n}\ge 0,~\forall i_1,i_2,\cdots, i_n, \\
        &\sum_{i_1\in\mathcal{X}_1}\sum_{i_2\in\mathcal{X}_2}\cdots\sum_{i_n\in\mathcal{X}_n}\tilde{\rho}^{A'_1A'_2\cdots A'_n}_{i_1i_2\cdots i_n}=\bigotimes_{j=1}^n I^{A'_j}  \\
         s.t.~~& P(i_1,i_2,\cdots ,i_n|\omega^{A'_1}_{k_1},\omega^{A'_2}_{k_2},\cdots ,\omega^{A'_n}_{k_n})\\
         &=\tr\left[\tilde{\rho}^{A'_1A'_2\cdots A'_n}_{i_1i_2\cdots i_n}\times\left(\bigotimes_{j=1}^n\omega_{k_j}^{A'_j}\right)\right]
        ,~~ \forall i_1,i_2,\cdots, i_n.
    \end{split}
\end{equation}

Compared to existing works on MDI entanglement quantification, we generalize the methods in \cite{PracticalQuant} to multipartite cases and do not require a case-by-case treatment of entanglement monotones as in \cite{Skrzypczyk2017}.

\section{MDI characterization of quantum memories}
\label{QuantumMemories}

In this section, we focus on the characterization of the resource theory of quantum memories given by Eq.~\eqref{QRTMEM}. Our framework on MDI resource characterization protocols is stated for static resource theories, but we can show that by exploiting the channel-state duality, the expectation-based and optimization-based protocols can still be applied with slight modifications. We also demonstrate that a novel resource characterization protocol based on the setting of key distribution exists for the resource theory of quantum memories.

\iffalse
In the context of MDI characterization of quantum memories, we regard entanglement-breaking (EB) channels as free objects and non-EB channels as resources. Following Definition 1, an MDI quantum memories verification protocol has the ability to successfully verify some non-EB channels if the measurements are faithful and will not wrongly conclude an EB channel to be non-EB due to imperfect detection. We shall present an MDI quantum memories verification protocol in Sec. \MakeUppercase{\romannumeral 5} A. And in Sec. \MakeUppercase{\romannumeral 5} B, we give an MDI quantification protocol taking the inspiration from the technique used in Sec. \MakeUppercase{\romannumeral 4} B.
\fi

\subsection{Expectation-based protocols}

As discussed earlier, we only need to focus on the MDI verification of the corresponding Choi state's entanglement.

Given a quantum memory, which can be described by a quantum channel $\mathcal{N}^{A\to B}\in\mathcal{Q}(\mathcal{H}^A\to\mathcal{H}^B)$. Similar to Theorem \ref{MDIinGeneral}, we have the following theorem that describes the existence of the MDI characterization for quantum memories, which relies on the characterization of bipartite entanglement.

\begin{theorem}[Existence of MDI protocols for characterization of quantum memories]
For any expectation-based $(\mathcal{R}_{\text{BI}},c,0,m)$-resource characterization protocol, where $\mathcal{R}_{\text{BI}}=(\mathcal{D}(\mathcal{H}^{A}\otimes\mathcal{H}^B),\text{SEP},\text{LOCC},\mu_R)$, there exists a $(\mathcal{R}_{\text{MEM}},c',0,m)$-resource characterization protocol with $c'\le c$. 
\end{theorem}

\begin{figure}[!htbp]
\label{FigDyanmicEnt}
	\includegraphics[scale=0.2]{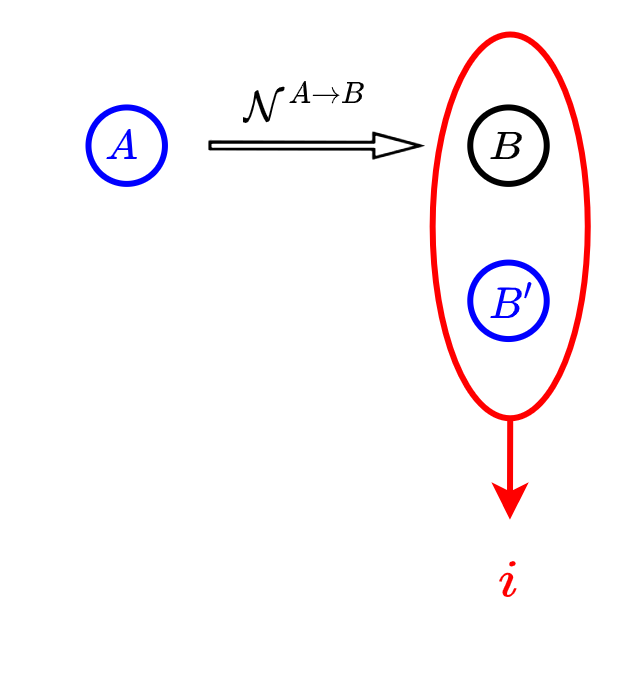}
    \caption{The setting for the MDI characterization of quantum memories. Here, the input trusted state $\rho^A$ undergoes the action of the memory $\mathcal{N}^{A\to B}$, and the outcome of this state is jointly measured with another trusted state.}
\end{figure}

\textit{Proof.} We first notice that $\mu_R$, as a resource monotone, is obviously invariant under transposition. Then, similar to Box \ref{BoxConversion}, we provide a constructive workflow in Box \ref{BoxMemory}, with a pictorial illustration in Fig. \ref{FigDyanmicEnt}.

\begin{mybox}[label={BoxMemory}]{{Constructing MDI protcols for $\mathcal{R}_{\text{MEM}}$ from protocols for $\mathcal{R}_{\text{BI}}$}}
\begin{enumerate}[label=(\Roman*)]
\item Draw an observable $\mathcal{W}$ from the expectation-based $(\mathcal{R}_{\text{BI}},c,0,m)$-resource characterization protocol.
\item Apply the local state decomposition and obtain the form
\begin{equation}
\mathcal{W}=\sum_{s=0}^{d^2_A-1}\sum_{t=0}^{d^2_B-1}\beta^i_{st}{\tau^T_s}\otimes(V_i{\omega^T_t}V^\dagger_i),
\end{equation}
where $V_i$ is a unitary transformation with respect to the generalized Bell states on $\mathcal{H}^{B'}$, and $\{\tau^{A}_s\}$ and $\{\omega^{B'}_t\}$ are two sets of local states on subsystems $A$ and $B'$.
\item Prepare the set of states $\tau^A_s$ using a trusted source and send them through the channel $\mathcal{N}^{A\to B}$.
\item Prepare another set of states $\omega^{B'}_t$ using a trusted source and perform a generalized Bell measurement on $\mathcal{N}(\tau^A_s)\otimes\omega^{B'}_t$. The measurement yields a set of probability distributions
\begin{equation}
    \left\{P(i|\tau_s^A,\omega_t^{B'})\right\}^i_{s,t}.
\end{equation}
\item Evaluate the MDI value
\begin{equation}
\mathcal{I}(\mathcal{N})=\sum_{s=0}^{d^2_A-1}\sum_{\substack{t=0 \\ i=0 }}^{d^2_B-1}\beta^i_{st}P(i|\tau_s^A,\omega_t^{B'}).
\end{equation}
\item The remaining steps are the same as in Box \ref{BoxConversion}.
\end{enumerate} 
\end{mybox}

\begin{figure}[!htbp]
	\includegraphics[scale=0.2]{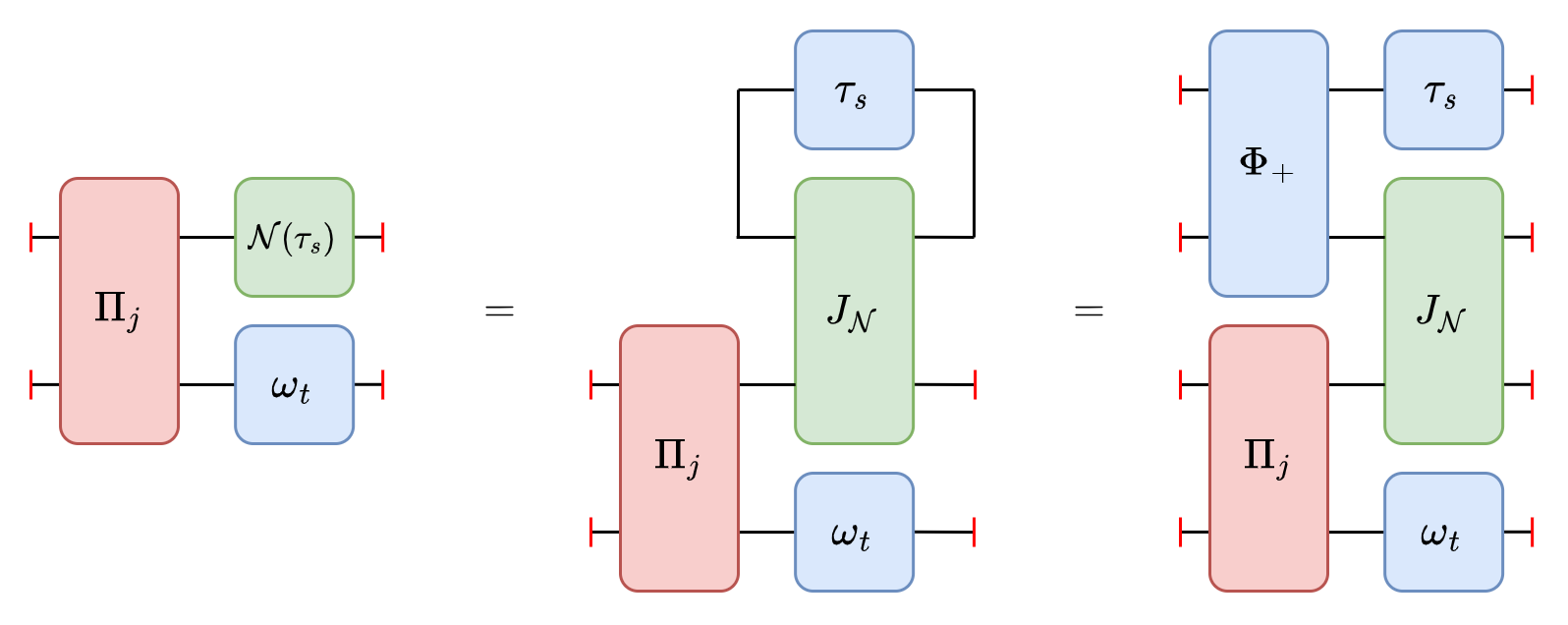}
    \caption{Illustration of how the characterization of quantum memories is equivalent to the characterization of Choi state bipartite entanglement, with one of the two measurements fixed to be $\Phi_+$. Here, we omit the normalization coefficients.}
    \label{FigMemoryEquivalence}
\end{figure}

The rest of the proof follows Theorem \ref{MDIinGeneral}, where the number of parties is restricted to $n=2$. The only difference is that we only require one generalized Bell measurement. This is due to the fact that 
\begin{equation}
\begin{aligned}
    \mathcal{N}^{A\to B}(\tau_s^A)&=d_A\tr_A\left[J_\mathcal{N}^{AB}(\tau_s^A)^T\otimes I^B\right]\\
    &=d_A^2\tr_{AA'}\left[(\Phi_+^{AA'}\otimes I^B)\times (\tau^{A'}_s\otimes J^{AB}_\mathcal{N})\right].
\end{aligned}
\end{equation}
Therefore, we have effectively performed a projection onto $\Phi_+$ in subsystem $AA'$. This observation is schematically shown in Fig. \ref{FigMemoryEquivalence}. $\square$

\subsection{Optimization-based protocols}

Similar to the discussion in Section \ref{SectionOpt}, we can also formulate the optimization problem for characterizing quantum memories:

\begin{equation}
    \begin{split}
        \min~~& \mu_R(J_\mathcal{N}) \\
        over~~&  \Pi_i\geq 0,~\forall
        i,~~\sum_i\Pi_i=I^B\otimes I^{B'}, \\
        & \mathcal{N}\in\mathcal{Q}(\mathcal{H}^A\to\mathcal{H}^B) \\
         s.t.~~& \tr [\Pi_i(\mathcal{N}(\tau_s)\otimes\omega_t)]=P(i|\tau_s,\omega_t),~~\forall i,s,t.
    \end{split}
\end{equation}

One can also relax the optimization problem into an SDP program:
\begin{equation}
    \begin{split}
        \min~~& \frac{1}{d_Ad_B}\sum_i\mu_R(\tilde{\rho}^{A'B'}_i) \\
        over~~&  \tilde{\rho}^{A'B'}_i\geq 0,~\forall
        i,~~\sum_i\tilde{\rho}^{A'B'}_i\leq I^{A'}\otimes I^{B'} \\
         s.t.~~& \tr [\tilde{\rho}^{A'B'}_i(\tau_s\otimes\omega_t)]=P(i|\tau_s,\omega_t),~~\forall i,s,t.
    \end{split}
\end{equation}
Where 
\begin{equation}
\tilde{\rho}^{A'B'}_i=d_A^2\tr_{AB}\left[\left(\Phi_+^{AA'}\otimes \Pi^{BB'}_i\right)\times \left(I^{A'}\otimes J^{AB}_\mathcal{N}\otimes  I^{B'}\right)\right],
\end{equation}
and the second inequality comes from the fact that $\Phi_+^{AA'}$ is only one of the possible POVM elements, thus it does not constitute a full set of POVM.

\subsection{Key-distribution-based protocols}

Here, we will show that a variant of Quantum Key Distribution (QKD) can be viewed as an MDI resource characterization protocol. Consider two parties, Alice and Bob, who share identical copies of the bipartite state $\rho^{AB}$, while an eavesdropper, Eve, holds their purification. Alice and Bob try to obtain as many correlated bits that are unknown to Eve as possible through operations that belong to LOCC. This paradigm falls within the well-studied field of QKD \cite{QKDReview}. It is known that the maximal distillable key rate $K$ is an operational measure of entanglement \cite{EntReview}, as it cannot be further increased under LOCC operations, and separable states are unable to distill any secret key. It is also straightforward to verify that $K$ is invariant under transposition, because the bit and phase error rates are not affected by sharing a globally transposed quantum state.

Although conventional (entanglement-based) QKD protocols can be used to characterize entanglement, they are vulnerable to attacks on the measurement end \cite{TimeShift}. A solution called MDI-QKD \cite{MDIQKD} has been proposed to close all measurement loopholes. 

We summarize the protocol in Box \ref{BoxMDIQKD}, with a slight modification for simplicity. In the original protocol, Alice and Bob send their quantum states to an untrusted third party, Charles. Therefore, two quantum communication channels, $\mathcal{N}^{A\to C}$ and $\mathcal{N}^{B\to C}$, are involved. In our modified version, we involve only one communication channel and set the untrusted third party as the untrusted measurement by Bob. We also generalize the intended measurement to include any type of joint measurement, where the original protocol focuses on a Bell state measurement in the context of quantum optical experiments. Lastly, we restrict ourselves to the qubit case, where there is only one type of phase error.

\begin{mybox}[label={BoxMDIQKD}]{{The MDI-QKD protocol}}
\begin{enumerate}[label=(\Roman*)]
\item Alice and Bob each randomly and independently prepare one of the four states: $\ket{0},\ket{1},\ket{+},\ket{-}$.
\item Alice sends her qubit state to Bob through a communication channel $\mathcal{N}^{A\to B}$.
\item Bob performs an uncharacterized joint measurement, described by a POVM $\Pi^{BB'}\in\mathscr{M}(\mathcal{H}^{BB'};\mathcal{X})$, on Alice's transmitted qubit and his own locally prepared qubit, where $\mathcal{X}=\{0,1,2,3\}$.
\item When Bob obtains an outcome $a\in\mathcal{X}$, he announces that his measurement event is successful, through a classical public channel.
\item Alice and Bob keep the data that corresponds to Bob's successful measurement events and discard the rest. 
\item Alice and Bob estimate the bit and phase error rates and perform key distillation, a process of classical post-processing, to obtain a secret key. Denote the maximal distillable key rate in this case as $K_a$.
\end{enumerate}
\end{mybox}
In the last step, denote the probability obtained by Alice and Bob as $p(a|\ket{\psi},\ket{\phi})$, representing the probability for a successful event to occur when Alice inputs $\ket{\psi}$ and Bob inputs $\ket{\phi}$. Then the bit error rate $e_b$ and phase error rate $e_p$ can be expressed by

\begin{equation}
    \begin{split}
        e_b &= \frac{\tr(\ketbra{0}{0}\otimes\ketbra{1}{1}\tilde{\rho}^T_a)+\tr(\ketbra{1}{1}\otimes\ketbra{0}{0}\tilde{\rho}^T_a)}{\tr\tilde{\rho}^T_a} \\
        &= \frac{p(a|\ket{0},\ket{1})+p(a|\ket{1},\ket{0})}{p(a|\ket{0},\ket{0})+p(a|\ket{0},\ket{1})+p(a|\ket{1},\ket{0})+p(a|\ket{1},\ket{1})},
    \end{split}
\end{equation}
\begin{equation}
    \begin{split}
        e_p &= \frac{\tr(\ketbra{+}{+}\otimes\ketbra{-}{-}\tilde{\rho}^T_a)+\tr(\ketbra{-}{-}\otimes\ketbra{+}{+}\tilde{\rho}^T_a)}{\tr\tilde{\rho}^T_a} \\
        &= \frac{p(a|\ket{+},\ket{-})+p(a|\ket{-},\ket{+})}{p(a|\ket{+},\ket{+})+p(a|\ket{+},\ket{-})+p(a|\ket{-},\ket{+})+p(a|\ket{-},\ket{-})},
    \end{split}
\end{equation}
where, according to the relationship between the BB84 protocol \cite{BB84} and the BBM92 protocol \cite{BBM92}, $\tilde{\rho}^T_a$ is the equivalent unnormalized state that Alice and Bob share when a successful event occurs:
\begin{equation}
    \tilde{\rho}_a=4\tr_{AB}\left[(\Phi^{AA'}_+\otimes \Pi_a^{BB'})\times\left(I^{A'}\otimes J_\mathcal{N}^{AB}\otimes I^{B'}\right)\right].
\end{equation}

We can give an MDI lower bound of the quantumness contained in the channel $\Lambda$:
\begin{equation}
    \begin{split}
    K(J_\mathcal{N}) &\geq \frac{1}{4}\sum_{j\in\mathcal{X}} K(\tilde{\rho}_j)\geq \frac{1}{4}K(\tilde{\rho}^T_a) \\
    &=\frac{1}{4}K_a(p(a|\ket{0},\ket{0})+p(a|\ket{0},\ket{1})+p(a|\ket{1},\ket{0})+p(a|\ket{1},\ket{1})) \\
    &=\frac{1}{4}K_a(p(a|\ket{+},\ket{+})+p(a|\ket{+},\ket{-})+p(a|\ket{-},\ket{+})+p(a|\ket{-},\ket{-})),
    \end{split}
\end{equation}
where the first inequality comes from Lemma \ref{POVMBound}, and $K_a$ can be calculated using $e_b$ and $e_p$ obtained.

Therefore, we can conclude that, contrary to QKD protocols based on entanglement distribution, which serve as resource characterization protocols for static entanglement, MDI-QKD protocols characterize the quantumness of the communication channel involved.

\section{Conclusion and outlook}

In this work, we introduced the concept of MDI resource characterization protocols and provided a general framework for them. We demonstrated the information-theoretical security of MDI protocols against any type of measurement imperfections. We also proved the existence of MDI protocols for LOSR-free resources and provided a general procedure to transform any expectation-based characterization protocol into an MDI version. As an example, we applied our framework to bipartite and multipartite entanglement, as well as non-EB quantum channels. Since LOSR operations are a natural free operation in the resource theory of nonlocality, this work bridges the gap between two subjects: resource theories of quantum nonlocality and MDI quantum information processing. By using our framework, we generalized, unified, and simplified many existing MDI protocols and proposed new ones, including the MDI characterization of multipartite entanglement classification, the MDI quantification of channel quantumness via SDP and key distribution. We also provided MDI protocols for nonlinear entanglement witnessing that do not require quantum memories or post-processing, making the protocol more applicable to near-term experiments while at the same time avoiding loopholes introduced by imperfect measurement devices.

There are still several aspects of MDI quantum information processing that are not covered by our framework. One example is the continuous variable MDI entanglement witness \cite{CVMDIEW}, where the quadrature formalism exhibits new structures that are absent in finite-dimensional Hilbert spaces. Another example is MDI quantum random number generation (MDI-QRNG) \cite{Cao_2015}, whose performance relies on measurement tomography rather than a specific form of nonlocality. It would also be interesting to further explore the relationship between MDI and DI resource characterization protocols, namely in what sense can we relax the trust on source devices. Moreover, the single shot regime for MDI resource characterization protocols remains less explored. For instance, it is possible to discuss entanglement testing \cite{berta_tangled_2024} within our framework, which aims to perform hypothesis testing of a potentially entangled state by measuring finite copies. We will leave these topics for future investigations.

\section*{Acknowledgement}

The authors thank Xiao Yuan and Ryuji Takagi for insightful discussions and comments. This work was supported by
the National Natural Science Foundation of China Grants No. 12174216 and the Innovation Program for Quantum
Science and Technology Grant No. 2021ZD0300804 and No. 2021ZD0300702.

\appendix

\section{Proof of Proposition \ref{LOSRentanglement}}
\label{pfLOSRentanglement}

It suffices to show that
\begin{equation}
    \mathcal{E}(\mathcal{F}_{\left\{\mathfrak{A}_1\right\}\left\{\mathfrak{A}_2\right\}\cdots\left\{\mathfrak{A}_k\right\}})\subseteq \mathcal{F}_{\left\{\mathfrak{A}_1\right\}\left\{\mathfrak{A}_2\right\}\cdots\left\{\mathfrak{A}_k\right\}}
\end{equation}
for any partition $\left\{\mathfrak{A}_1\right\}\left\{\mathfrak{A}_2\right\}\cdots\left\{\mathfrak{A}_k\right\}$ of $\mathcal{H}$.

The LOSR operation $\mathcal{E}$ has the form
\begin{equation}
\begin{aligned}
    \mathcal{E}&=\sum_\mu\pi(\mu)\bigotimes_{j=1}^n \Phi_\mu^{A_j}\\
    &=\sum_\mu\pi(\mu)\bigotimes_{j=1}^k \Phi_\mu^{\mathfrak{A}_j},
\end{aligned}
\end{equation}
where $\Phi_\mu^{A_j}\in\mathcal{Q}(\mathcal{H}^{A_j}\to\mathcal{H}^{A_j})$ and $\Phi_\mu^{\mathfrak{A}_j}\in\mathcal{Q}(\mathcal{H}^{\mathfrak{A}_j}\to\mathcal{H}^{\mathfrak{A}_j})$. The second equality follows from tensoring the local channels with respect to the partition $\left\{\mathfrak{A}_1\right\}\left\{\mathfrak{A}_2\right\}\cdots\left\{\mathfrak{A}_k\right\}$.

Since the action of $\bigotimes_{j=1}^k\Phi_\mu^{\mathfrak{A}_j}$ on $\rho\in\mathcal{F}_{\left\{\mathfrak{A}_1\right\}\left\{\mathfrak{A}_2\right\}\cdots\left\{\mathfrak{A}_k\right\}}$ is obviously another state in $\mathcal{F}_{\left\{\mathfrak{A}_1\right\}\left\{\mathfrak{A}_2\right\}\cdots\left\{\mathfrak{A}_k\right\}}$, $\mathcal{E}(\rho)$ is simply a convex combination of states in $\mathcal{F}_{\left\{\mathfrak{A}_1\right\}\left\{\mathfrak{A}_2\right\}\cdots\left\{\mathfrak{A}_k\right\}}$, which completes the proof.

\section{Proof of Lemma \ref{StateDecomp}}
\label{pfStateDecomp}

We provide a constructive proof of the lemma. Notice that there exists an operator basis $\{B_j\}$, $j=0,\cdots, d^2-1$ on $\mathcal{H}^d$ satisfying the following properties:
\begin{enumerate}
    \item $B_0=I_d,\tr B_j=0,\forall j =1,\cdots, d^2-1$.
    \item $\tr B_j^\dagger B_k=c \delta_{jk}$, where $c\in\mathbb{R}$. In other words, the basis is orthogonal under the Hilbert-Schmidt inner product. 
\end{enumerate}

Using the property $\tr(X\otimes Y)=\tr X\cdot \tr Y$, elements of the form $\bigotimes_{j=1}^n B_{k_j}^{A_j}, (k_j = 0,\cdots, d^2_{A_j}-1)$ are mutually orthogonal under the Hilbert-Schmidt inner product, and the number of these elements is equal to $\operatorname{dim}(\mathcal{B}(\mathcal{H}))$. Therefore, the elements form a complete basis of $\mathcal{B}(\mathcal{H})$:
\begin{equation}
    \mathcal{W}=\sum_{k_1=0}^{d_{A_1}^2-1}\sum_{k_2=0}^{d_{A_2}^2-1}\cdots \sum_{k_n=0}^{d_{A_n}^2-1}\gamma_{k_1k_2\cdots k_n}\bigotimes_{j=1}^n B_{k_j}^{A_j},
\end{equation} where $\gamma_{k_1k_2\cdots k_n}=\frac{1}{c^n}\tr({(\bigotimes_{j=1}^n B_{k_j}^{A_j})}^\dagger\mathcal{W})$.

In order to obtain a local state decomposition, we need to perform a substitution
\begin{equation}
    \tilde{\tau}_{k_j}^{A_j}=B_{k_j}^{A_j} + a_{k_j}\cdot B_0^{A_j},
\end{equation}
where $a_{k_j}$ is the smallest coefficient to make $\tilde{\tau}_{k_j}^{A_j}$ positive semi-definite. We add a tilde to indicate that it is an unnormalized quantum state. Since $\tilde{\tau}_{k_j}^{A_j}$ and $B_0^{A_j}$ are all positive semi-definite, after rearranging the terms in the expansion and a proper normalization, we obtain the form in Lemma \ref{StateDecomp}.

\section{Explicit construction of MDI value}
\label{explicit}

We provide an explicit construction of the MDI value Eq.~\eqref{MDIvalue}, or equivalently, the expansion coefficients used for classical post-processing of the obtained probabilities. For the standard operator basis, we choose the generalized Gell-Mann matrices \cite{QuditBloch}, stemming from standard $SU(d)$ Hermitian generators. The $d^2-1$ matrices can be categorized into $d(d-1)/2$ symmetric matrices, $d(d-1)/2$ antisymmetric matrices, and $d-1$ diagonal ones:
\begin{equation}
\begin{gathered}
\Lambda^s_{\mu\nu}=\ketbra{\mu}{\nu}+\ketbra{\nu}{\mu},\quad 0\le \mu<\nu\le d-1,\\
\Lambda^a_{\mu\nu}=-i\ketbra{\mu}{\nu}+i\ketbra{\nu}{\mu},\quad 0\le \mu<\nu\le d-1,\\
\Lambda_\lambda=\sqrt{\frac{2}{(\lambda+1)(\lambda+2)}}\left(\sum_{j=0}^\lambda \ketbra{j}{j}-(\lambda+1)\ketbra{\lambda+1}{\lambda+1}\right), \quad 0 \leq \lambda \leq d-2 .
\end{gathered}
\end{equation}
When $d=2$, the three cases correspond to Pauli operators $X,Y,$ and $Z,$ respectively. These traceless Hermitian matrices, together with the identity, constitute the operator basis on $\mathcal{H}^d$. To keep the notations compact, we merge the indices into a single index, namely $B_0=I$, $B_1\sim B_{d-1}$ are the $d-1$ matrices $\Lambda_\lambda$, $B_{d}\sim B_{\frac{1}{2}d(d+1)-1}$ are the $d(d-1)/2$ matrices $\Lambda^s_{\mu\nu}$, and finally $B_{\frac{1}{2}d(d+1)}\sim B_{d^2-1}$ are the $d(d-1)/2$ matrices $\Lambda^a_{\mu\nu}$.

The smallest shift to make the corresponding state positive semi-definite can be explicitly given by
\begin{equation}
    a_k = \left\{\begin{aligned}
    0&, \quad k = 0\\
    \sqrt{\frac{2k}{k+1}}&,\quad k = 1,2,\cdots, d-1\\
    1&,\quad k = d,d+1,\cdots, d^2-1.
    \end{aligned}\right.
\end{equation}
The normalized quantum states can therefore be represented by 
\begin{equation}
\tau_k = \left\{\begin{aligned}
    B_0/d&, \quad k = 0\\
    B_0/d + \sqrt{\frac{k+1}{2k}}B_k/d &,\quad k = 1,2,\cdots, d-1\\
    B_0/d + B_k/d &,\quad k = d,d+1,\cdots, d^2-1.
    \end{aligned}\right.
\end{equation}
To describe the generalized Bell states, we use the Heisenberg-Weyl basis:
\begin{equation}
    U_{n m}=\sum_{k=0}^{d-1} e^{\frac{2 \pi i}{d} k n}\ketbra{k}{(k+m) \bmod d} \quad n, m=0,1, \ldots, d-1.
\end{equation}
Again, to make the indices more compact, we denote $i = n\cdot d + m$ when necessary.

The action of $U_{nm}$ on the generalized Gell-Mann matrices can be calculated:
\begin{equation}
    \begin{aligned}
        U_{nm}\Lambda^s_{\mu\nu}U^\dagger_{nm} & = \sum_{k=0}^{d-1} e^{\frac{2 \pi i}{d} k n}\ketbra{k}{(k+m) \bmod d}\left[\ketbra{\mu}{\nu}+\ketbra{\nu}{\mu}\right]\sum_{k'=0}^{d-1} e^{-\frac{2 \pi i}{d} k' n}\ketbra{(k'+m) \bmod d}{k'}\\
        & = e^{\frac{2\pi i}{d}n(\mu-\nu)}\ketbra{\mu- m \bmod d}{\nu- m \bmod d} + e^{-\frac{2\pi i}{d}n(\mu-\nu)}\ketbra{\nu- m \bmod d}{\mu- m \bmod d},\\
        U_{nm}\Lambda^a_{\mu\nu}U^\dagger_{nm} & = \sum_{k=0}^{d-1} e^{\frac{2 \pi i}{d} k n}\ketbra{k}{(k+m) \bmod d}\left[-i\ketbra{\mu}{\nu}+i\ketbra{\nu}{\mu}\right]\sum_{k'=0}^{d-1} e^{-\frac{2 \pi i}{d} k' n}\ketbra{(k'+m) \bmod d}{k'}\\
        & = -ie^{\frac{2\pi i}{d}n(\mu-\nu)}\ketbra{\mu- m \bmod d}{\nu- m \bmod d} + ie^{-\frac{2\pi i}{d}n(\mu-\nu)}\ketbra{\nu- m \bmod d}{\mu- m \bmod d},\\
        U_{nm}\Lambda_\lambda U^\dagger_{nm} & = \sqrt{\frac{2}{(\lambda+1)(\lambda+2)}}\left(\sum_{j=0}^\lambda\ketbra{j-m\bmod d}{j-m\bmod d}-(\lambda+1)\ketbra{\lambda+1-m\bmod d}{\lambda+1-m\bmod d}\right).
    \end{aligned}
\end{equation}

By plugging in the expansion of matrix units $\ketbra{j}{k}$ using generalized Gell-Mann matrices, we have
\begin{equation}
    \ketbra{j}{k}= \begin{cases}\frac{1}{2}\left(\Lambda^s_{j k}+i \Lambda^a_{j k}\right) & \text { for } j<k \\ \frac{1}{2}\left(\Lambda^s_{k j}-i \Lambda^a_{k j}\right) & \text { for } j>k \\ -\sqrt{\frac{j}{2 (j+1)}} \Lambda_{j-1}+\sum_{n=0}^{d-j-2} \frac{1}{\sqrt{2(j+n+1)(j+n+2)}} \Lambda_{j+n}+\frac{1}{d} I & \text { for } j=k.\end{cases}
\end{equation}
We can express the new basis $U_i\tau_k U_i^\dagger$ using the original Hermitian operator basis $\{B_k\}$. In other words, we can now directly calculate the transformation between $U_i\tau_k U_i^\dagger$ and $\{B_k\}$, differing from the protocol in the main text, which requires two steps.

In conclusion, the coefficients for the MDI value can be explicitly expressed as
\begin{equation}
    \begin{aligned}
        \beta_{k_1k_2\cdots k_n}^{i_1i_2\cdots i_n} =\frac{1}{2^n}\sum_{k'_1=0}^{d_{A_1}^2-1}\sum_{k'_2=0}^{d_{A_2}^2-1}\cdots \sum_{k'_n=0}^{d_{A_n}^2-1}\left[\prod_{j=1}^{n}\left(\mathcal{T}^{-1}_{i_j}\right)^{A_j}_{k'_jk_j}\right]\tr\left(\mathcal{W}\bigotimes_{j=1}^n B^{A_j}_{k'_j}\right),
    \end{aligned}
\end{equation}
where $\mathcal{T}_i$ is the transformation matrix between ${\{U_i\tau_kU_i^\dagger\}}_k$ and ${\{B_k\}}_k$: $U_i\tau_kU_i^\dagger=\sum_{k'}(\mathcal{T}_i)_{kk'}B_{k'}$.

The closed form for the general expression of $\mathcal{T}_i$ and its inversion is rather complicated. Here, we take the case of two qutrits as an example and calculate the explicit form of $\mathcal{T}_{nm}$, which satisfies $U_{nm}\tau_kU_{nm}^\dagger=\sum_{k'}(\mathcal{T} _{nm})_{kk'}B_{k'}$. Each $\mathcal{T}$ matrix is a $9\times9$ matrix:
\begin{equation}
\mathcal{T}_{n0} =\frac{1}{3} \begin{pmatrix}
    1&0&0&0&0&0&0&0&0 \\
    1&1&0&0&0&0&0&0&0 \\
    1&0&\frac{\sqrt{3}}{2}&0&0&0&0&0&0 \\
    1&0&0&\cos(\frac{2}{3}\pi n)&0&0&\sin(\frac{2}{3}\pi n)&0&0\\
    1&0&0&0&\cos(\frac{4}{3}\pi n)&0&0&\sin(\frac{4}{3}\pi n)&0 \\
    1&0&0&0&0&\cos(\frac{2}{3}\pi n)&0&0&\sin(\frac{2}{3}\pi n) \\
    1&0&0&-\sin(\frac{2}{3}\pi n)&0&0&\cos(\frac{2}{3}\pi n)&0&0 \\
    1&0&0&0&-\sin(\frac{4}{3}\pi n)&0&0&\cos(\frac{4}{3}\pi n)&0\\
    1&0&0&0&0&-\sin(\frac{2}{3}\pi n)&0&0&\cos(\frac{2}{3}\pi n) \\
\end{pmatrix},
\end{equation}

\begin{equation}
\mathcal{T}_{n1} =\frac{1}{3} \begin{pmatrix}
    1&0&0&0&0&0&0&0&0 \\
    1&-\frac{1}{2}&-\frac{\sqrt{3}}{2}&0&0&0&0&0&0 \\
    1&\frac{3}{4}&-\frac{\sqrt{3}}{4}&0&0&0&0&0&0 \\
    1&0&0&0&\cos(\frac{2}{3}\pi n)&0&0&-\sin(\frac{2}{3}\pi n)&0\\
    1&0&0&0&0&\cos(\frac{4}{3}\pi n)&0&0&-\sin(\frac{4}{3}\pi n) \\
    1&0&0&\cos(\frac{2}{3}\pi n)&0&0&\sin(\frac{2}{3}\pi n)&0&0 \\
    1&0&0&0&-\sin(\frac{2}{3}\pi n)&0&0&-\cos(\frac{2}{3}\pi n)&0 \\
    1&0&0&0&0&-\sin(\frac{4}{3}\pi n)&0&0&-\cos(\frac{4}{3}\pi n) \\
    1&0&0&-\sin(\frac{2}{3}\pi n)&0&0&\cos(\frac{2}{3}\pi n)&0&0 \\
\end{pmatrix},
\end{equation}

\begin{equation}
\mathcal{T}_{n2} =\frac{1}{3} \begin{pmatrix}
    1&0&0&0&0&0&0&0&0 \\
    1&-\frac{1}{2}&\frac{\sqrt{3}}{2}&0&0&0&0&0&0 \\
    1&-\frac{3}{4}&-\frac{\sqrt{3}}{4}&0&0&0&0&0&0 \\
    1&0&0&0&0&\cos(\frac{2}{3}\pi n)&0&0&\sin(\frac{2}{3}\pi n) \\
    1&0&0&\cos(\frac{4}{3}\pi n)&0&0&-\sin(\frac{4}{3}\pi n)&0&0 \\
    1&0&0&0&\cos(\frac{2}{3}\pi n)&0&0&-\sin(\frac{2}{3}\pi n)&0 \\
    1&0&0&0&0&-\sin(\frac{2}{3}\pi n)&0&0&\cos(\frac{2}{3}\pi n) \\
    1&0&0&-\sin(\frac{4}{3}\pi n)&0&0&-\cos(\frac{4}{3}\pi n)&0&0 \\
    1&0&0&0&-\sin(\frac{2}{3}\pi n)&0&0&-\cos(\frac{2}{3}\pi n)&0 \\
\end{pmatrix}.
\end{equation}
The inverses are given by:
\begin{small}
\begin{equation}
\mathcal{T}^{-1}_{n0} =3 \begin{pmatrix}
    1&0&0&0&0&0&0&0&0 \\
    -1&1&0&0&0&0&0&0&0 \\
    -\frac{2}{\sqrt{3}}&0&\frac{2}{\sqrt{3}}&0&0&0&0&0&0 \\
    -\cos(\frac{2}{3}\pi n)+\sin(\frac{2}{3}\pi n)&0&0&\cos(\frac{2}{3}\pi n)&0&0&-\sin(\frac{2}{3}\pi n)&0&0\\
    -\cos(\frac{4}{3}\pi n)+\sin(\frac{4}{3}\pi n)&0&0&0&\cos(\frac{4}{3}\pi n)&0&0&-\sin(\frac{4}{3}\pi n)&0 \\
    -\cos(\frac{2}{3}\pi n)+\sin(\frac{2}{3}\pi n)&0&0&0&0&\cos(\frac{2}{3}\pi n)&0&0&-\sin(\frac{2}{3}\pi n) \\
    -\cos(\frac{2}{3}\pi n)-\sin(\frac{2}{3}\pi n)&0&0&\sin(\frac{2}{3}\pi n)&0&0&\cos(\frac{2}{3}\pi n)&0&0 \\
    -\cos(\frac{4}{3}\pi n)-\sin(\frac{4}{3}\pi n)&0&0&0&\sin(\frac{4}{3}\pi n)&0&0&\cos(\frac{4}{3}\pi n)&0\\
    -\cos(\frac{2}{3}\pi n)-\sin(\frac{2}{3}\pi n)&0&0&0&0&\sin(\frac{2}{3}\pi n)&0&0&\cos(\frac{2}{3}\pi n) \\
\end{pmatrix},
\end{equation}

\begin{equation}
\mathcal{T}^{-1}_{n1} =3 \begin{pmatrix}
    1&0&0&0&0&0&0&0&0 \\
    -\frac{1}{2}&-\frac{1}{2}&1&0&0&0&0&0&0 \\
    \frac{5}{2\sqrt{3}}&-\frac{\sqrt{3}}{2}&-\frac{1}{\sqrt{3}}&0&0&0&0&0&0 \\
    -\cos(\frac{2}{3}\pi n)+\sin(\frac{2}{3}\pi n)&0&0&0&0&\cos(\frac{2}{3}\pi n)&0&0&-\sin(\frac{2}{3}\pi n)\\
    -\cos(\frac{2}{3}\pi n)+\sin(\frac{2}{3}\pi n)&0&0&\cos(\frac{2}{3}\pi n)&0&0&-\sin(\frac{2}{3}\pi n)&0&0 \\
    -\cos(\frac{4}{3}\pi n)+\sin(\frac{4}{3}\pi n)&0&0&0&\cos(\frac{4}{3}\pi n)&0&0&-\sin(\frac{4}{3}\pi n)&0 \\
    -\cos(\frac{2}{3}\pi n)-\sin(\frac{2}{3}\pi n)&0&0&0&0&\sin(\frac{2}{3}\pi n)&0&0&\cos(\frac{2}{3}\pi n) \\
    \cos(\frac{2}{3}\pi n)+\sin(\frac{2}{3}\pi n)&0&0&-\sin(\frac{2}{3}\pi n)&0&0&-\cos(\frac{2}{3}\pi n)&0&0 \\
    \cos(\frac{4}{3}\pi n)+\sin(\frac{4}{3}\pi n)&0&0&0&-\sin(\frac{4}{3}\pi n)&0&0&-\cos(\frac{4}{3}\pi n)&0 \\
\end{pmatrix},
\end{equation}

\begin{equation}
\mathcal{T}^{-1}_{n2} =3 \begin{pmatrix}
    1&0&0&0&0&0&0&0&0 \\
    \frac{3}{2}&-\frac{1}{2}&-1&0&0&0&0&0&0 \\
    -\frac{1}{2\sqrt{3}}&\frac{\sqrt{3}}{2}&-\frac{1}{\sqrt{3}}&0&0&0&0&0&0 \\
    -\cos(\frac{4}{3}\pi n)+\sin(\frac{4}{3}\pi n)&0&0&0&\cos(\frac{4}{3}\pi n)&0&0&-\sin(\frac{4}{3}\pi n)&0 \\
    -\cos(\frac{2}{3}\pi n)+\sin(\frac{2}{3}\pi n)&0&0&0&0&\cos(\frac{2}{3}\pi n)&0&0&-\sin(\frac{2}{3}\pi n) \\
    -\cos(\frac{2}{3}\pi n)+\sin(\frac{2}{3}\pi n)&0&0&\cos(\frac{2}{3}\pi n)&0&0&-\sin(\frac{2}{3}\pi n)&0&0 \\
    \cos(\frac{4}{3}\pi n)+\sin(\frac{4}{3}\pi n)&0&0&0&-\sin(\frac{4}{3}\pi n)&0&0&-\cos(\frac{4}{3}\pi n)&0 \\
    \cos(\frac{2}{3}\pi n)+\sin(\frac{2}{3}\pi n)&0&0&0&0&-\sin(\frac{2}{3}\pi n)&0&0&-\cos(\frac{2}{3}\pi n) \\
    -\cos(\frac{2}{3}\pi n)-\sin(\frac{2}{3}\pi n)&0&0&\sin(\frac{2}{3}\pi n)&0&0&\cos(\frac{2}{3}\pi n)&0&0 \\
\end{pmatrix}.
\end{equation}
\end{small}

\section{Proof of Theorem \ref{MDIinGeneral}}
\label{pfMDIinGeneral}

For simplicity, we only consider single copy observables in the original expectation-based protocol. For multicopy observables, as discussed in Section \ref{MulticopyWitness}, can always be regarded as classical processing of the single copy observable scenario.

We first show that the new protocol has a completeness parameter of at most $c$. Eve only needs to do nothing and let Alice faithfully implement all the generalized Bell state measurements involved in the protocol. For simplicity, we only consider the case where $\mathcal{W}$ acts on a single copy of $\rho$. In this case,

\begin{equation}
\begin{aligned}
    P(i_1,i_2,\cdots ,i_n|\omega^{A'_1}_{k_1},\omega^{A'_2}_{k_2},\cdots ,\omega^{A'_n}_{k_n})&=\tr\left\{\left[\bigotimes_{j=1}^n\left(I^{A'_j}\otimes U_{i_j}^{A_j}\right)\Phi_+^{A'_jA_j}\left(I^{A'_j}\otimes (U^{A_j}_{i_j})^\dagger\right)\right]\times\left[\rho^{A_1A_2\cdots A_n}\otimes \bigotimes_{j=1}^n\omega_{k_j}^{A'_j}\right]\right\}\\
    & = \frac{1}{\prod_{j=1}^n d_{A_j}}\tr\left[\left(\bigotimes_{j=1}^n U_{i_j}\left(\omega_{k_j}^{A_j}\right)^T U^\dagger_{i_j}\right)\rho^{A_1A_2\cdots A_n}\right].
\end{aligned}
\end{equation}
Recall that $\left(\omega_{k_j}^{A_j}\right)^T=\tau_{k_j}^{A_j}$, the corresponding MDI value becomes
\begin{equation}
    \begin{aligned}
        \mathcal{I}(\rho) &= \sum_{\substack{i_1=0 \\ k_1=0 }}^{d_{A_1}^2-1} \sum_{\substack{i_2=0 \\ k_2=0 }}^{d_{A_2}^2-1} \cdots \sum_{\substack{i_n=0 \\ k_n=0 }}^{d_{A_n}^2-1} \beta_{k_1k_2\cdots k_n}^{i_1i_2\cdots i_n} P(i_1,i_2,\cdots ,i_n|\omega^{A'_1}_{k_1},\omega^{A'_2}_{k_2},\cdots ,\omega^{A'_n}_{k_n})\\
        &= \frac{1}{\prod_{j=1}^n d_{A_j}}\tr\left[\sum_{\substack{i_1=0 \\ k_1=0 }}^{d_{A_1}^2-1} \sum_{\substack{i_2=0 \\ k_2=0 }}^{d_{A_2}^2-1} \cdots \sum_{\substack{i_n=0 \\ k_n=0 }}^{d_{A_n}^2-1} \beta_{k_1k_2\cdots k_n}^{i_1i_2\cdots i_n}\left(\bigotimes_{j=1}^n U_{i_j} \tau_{k_j}^{A_j}U^\dagger_{i_j}\right)\rho^{A_1A_2\cdots A_n}\right]\\
        &= \frac{1}{\Omega}\tr(\mathcal{W}\rho),
    \end{aligned}
\end{equation}
where the third equality is due to Eq.~\eqref{DecomposedStateWithSuperscript}. The generalization to $\mathcal{W}$ acting on $N$ identical copies of $\rho$ is straightforward. We then obtain the result:
\begin{equation}
    \begin{aligned}
    C_{\text{MDI}}(g)&=f\left[\Omega^{N_1}\mathcal{I}_1(\rho), \Omega^{N_2}\mathcal{I}_2(\rho),\cdots,\Omega^{N_m}\mathcal{I}_m(\rho)\right]\\
    &=f\left[\tr(\mathcal{W}_1\rho^{\otimes N_1}),\tr(\mathcal{W}_2\rho^{\otimes N_2}),\cdots\tr(\mathcal{W}_m\rho^{\otimes N_m})\right]\\
    &=C(g).
    \end{aligned}
\end{equation}
Therefore, by letting Alice faithfully implement her generalized Bell state measurements, this strategy of Eve will ensure that Alice obtains the same value $C(g)$ in the original trusted-device protocol. Since the original resource characterization protocol has a completeness parameter $c$, the new protocol has a completeness parameter $c'\le c$.

We now demonstrate that the new protocol has a soundness parameter of 0. When Eve manipulates the measurement and implements the POVM element $\sum_\mu\pi(\mu)\bigotimes_{j=1}^n \Xi^{A_jA'_j}_{\mu}$ instead of the intended generalized Bell state measurement $\bigotimes_{j=1}^n \Pi^{A_jA'_j}$, we can obtain
\begin{equation}
\begin{aligned}
    P_{\text{Eve}}(i_1,i_2,\cdots ,i_n|\omega^{A'_1}_{k_1},\omega^{A'_2}_{k_2},\cdots ,\omega^{A'_n}_{k_n})&=\tr\left[\left(\sum_\mu\pi(\mu)\bigotimes_{j=1}^n \Xi^{A_jA'_j}_{\mu;i_j}\right)\times\left(\rho^{A_1A_2\cdots A_n}\otimes \bigotimes_{j=1}^n\omega_{k_j}^{A'_j}\right)\right]\\
    &=\tr\left(\sum_\mu\pi(\mu)\mathcal{N}_{\mu;i_1i_2\cdots i_n}^{A'_1A'_2\cdots A'_n}\bigotimes_{j=1}^{n}\omega_{k_j}^{A'_j}\right),
\end{aligned}
\end{equation}
where $\mathcal{N}^{A'_1A'_2\cdots A'_n}_{\mu;i_1i_2\cdots i_n}$ is defined by
\begin{equation}
    \mathcal{N}^{A'_1 A'_2\cdots A'_n}_{\mu;i_1i_2\cdots i_n}=\tr_{A_1A_2\cdots A_n}\left[\left(\bigotimes_{j=1}^n \Xi^{A_jA'_j}_{\mu;i_j}\right)\times\left(\bigotimes_{j=1}^nI^{A_j'}\otimes\rho^{A_1A_2\cdots A_n}\right)\right].
\end{equation}

One can verify that
\begin{equation}
    \mathcal{N}^{A'_1A'_2\cdots A'_n}_\mu\in\mathscr{M}\left(\bigotimes_{j=1}^n\mathcal{H}^{A'_j};\mathcal{X}_1\times\mathcal{X}_2\times\cdots\mathcal{X}_n\right)
\end{equation}
with each $\mathcal{X}_j$ being $\{0,1,\cdots,d_{A_j}^2-1\}$.

The corresponding MDI value in this case is given by
\begin{equation}
\begin{aligned}
    \mathcal{I}_{\text{Eve}}(\rho)&=\sum_{\substack{i_1=0 \\ k_1=0 }}^{d_{A_1}^2-1} \sum_{\substack{i_2=0 \\ k_2=0 }}^{d_{A_2}^2-1} \cdots \sum_{\substack{i_n=0 \\ k_n=0 }}^{d_{A_n}^2-1} \beta_{k_1k_2\cdots k_n}^{i_1i_2\cdots i_n} P_{\text{Eve}}(i_1,i_2,\cdots ,i_n|\omega^{A'_1}_{k_1},\omega^{A'_2}_{k_2},\cdots ,\omega^{A'_n}_{k_n})\\
    &=\sum_{\substack{i_1=0 \\ k_1=0 }}^{d_{A_1}^2-1} \sum_{\substack{i_2=0 \\ k_2=0 }}^{d_{A_2}^2-1} \cdots \sum_{\substack{i_n=0 \\ k_n=0 }}^{d_{A_n}^2-1} \beta_{k_1k_2\cdots k_n}^{i_1i_2\cdots i_n}\tr\left(\sum_\mu\pi(\mu)\mathcal{N}^{A'_1A'_2\cdots A'_n}_{\mu;i_1i_2\cdots i_n} \bigotimes_{j=1}^n\omega_{k_j}^{A'_j}\right)\\
    &=\sum_{\substack{i_1=0 \\ k_1=0 }}^{d_{A_1}^2-1} \sum_{\substack{i_2=0 \\ k_2=0 }}^{d_{A_2}^2-1} \cdots \sum_{\substack{i_n=0 \\ k_n=0 }}^{d_{A_n}^2-1} \beta_{k_1k_2\cdots k_n}^{i_1i_2\cdots i_n} \tr\left[\sum_{\mu}\pi(\mu)\left(\bigotimes_{j=1}^n U_{i_j}^*\right)\mathcal{N}^{A'_1A'_2\cdots A'_n}_{\mu;i_1i_2\cdots i_n}\left(\bigotimes_{j=1}^n U_{i_j}^T\right)\times \bigotimes_{j=1}^n\left(U_{i_j}^*\omega_{k_j}^{A'_j}U^T_{i_j}\right)\right]\\
    &=\sum_{i_1=0 }^{d_{A_1}^2-1} \sum_{i_2=0 }^{d_{A_2}^2-1} \cdots \sum_{ i_n=0 }^{d_{A_n}^2-1}\sum_\mu\pi(\mu)\tr\left[\left(\bigotimes_{j=1}^n U_{i_j}^*\right)\mathcal{N}^{A'_1A'_2\cdots A'_n}_{i_1i_2\cdots i_n}\left(\bigotimes_{j=1}^n U_{i_j}^T\right)\times \mathcal{W}^T\right]\\&=\tr\left\{\mathcal{W}\times\left[\sum_\mu\pi(\mu)\sum_{i_1=0 }^{d_{A_1}^2-1} \sum_{i_2=0 }^{d_{A_2}^2-1} \cdots \sum_{ i_n=0 }^{d_{A_n}^2-1}\left(\bigotimes_{j=1}^n U_{i_j}\right)\left(\mathcal{N}^{A'_1A'_2\cdots A'_n}_{\mu;i_1i_2\cdots i_n}\right)^T\left(\bigotimes_{j=1}^n U_{i_j}^\dagger\right)\right]\right\}\\
    &=\tr\left(\mathcal{W}\sigma^T_{\text{Eve}}\right),
\label{EveMDIValue}
\end{aligned}
\end{equation}
where $\sigma_{Eve}\in\mathcal{D}(\bigotimes_{j=1}^n\mathcal{H}^{A'_j})$ is a quantum state defined by
\begin{equation}
\begin{aligned}
    \sigma_{\text{Eve}}&=\sum_\mu\pi(\mu)\sum_{i_1=0 }^{d_{A_1}^2-1} \sum_{i_2=0 }^{d_{A_2}^2-1} \cdots \sum_{ i_n=0 }^{d_{A_n}^2-1}\left(\bigotimes_{j=1}^n U_{i_j}^*\right)\mathcal{N}^{A'_1A'_2\cdots A'_n}_{\mu;i_1i_2\cdots i_n}\left(\bigotimes_{j=1}^n U_{i_j}^T\right)\\
    &=\sum_\mu\pi(\mu)\sum_{i_1=0 }^{d_{A_1}^2-1} \sum_{i_2=0 }^{d_{A_2}^2-1} \cdots \sum_{ i_n=0 }^{d_{A_n}^2-1}\left(\bigotimes_{j=1}^n U_{i_j}^*\right)\tr_{A_1A_2\cdots A_n}\left[\left(\bigotimes_{j=1}^n \Xi^{A_jA'_j}_{\mu;i_j}\right)\times\left(\bigotimes_{j=1}^nI^{A_j'}\otimes\rho^{A_1A_2\cdots A_n}\right)\right]\left(\bigotimes_{j=1}^n U_{i_j}^T\right).
\end{aligned}
\end{equation}
$\sigma_{\text{Eve}}$ can be prepared by applying LOSR operations on $\rho$. We have the relation \begin{equation}
\mu(\sigma_{\text{Eve}}^T)=\mu(\sigma_\text{Eve})\le \mu(\rho)\le m.
\end{equation}

Since the original protocol has a soundness parameter of 0, it follows that the new protocol will reject $\sigma_{\text{Eve}}^T$ with certainty, as long as the original protocol rejects $\rho$.

\section{MDI protocols that allow classical communication}
\label{pfLOCC}
Here we discuss the situation where the experiment is not conducted in a spacelike separated scenario and communication between subsystems is allowed, in the context of entanglement characterization. In this case, instead of implementing the measurement in Eq.~\eqref{EvePOVM}, Eve can perform arbitrary adaptive measurement strategies and perform a so called ``LOCC measurement". A precise mathematical characterization of LOCC measurements is rather cumbersome. Therefore, we will focus on separable measurements \cite{SeparableOperation}, denoted as SEP, with the strict inclusion relation
\begin{equation}
    \text{LOCC}\subset \text{SEP},
\end{equation}
separable measurements have a much simpler structure. Namely, a separable measurement can be described by a POVM 
\begin{equation}
    \Xi\in\mathscr{M}(\bigotimes_{j=1}^{n}\mathcal{H}^{A_j};\mathcal{X}_1\times\mathcal{X}_2\times\cdots\times\mathcal{X}_n)
\end{equation}
with every POVM element $\Xi_{i_1i_2\cdots i_n}$ taking the form 
\begin{equation}
    \Xi_{i_1i_2\cdots i_n}=\sum_{k}\bigotimes_{j=1}^n M^{A_j}_{k;i_1i_2\cdots i_n},
\end{equation}
where $M^{A_j}_{k;i_1i_2\cdots i_n}$ are positive operators acting on $A_j$. We can then similarly repeat the proof of Theorem \ref{MDIinGeneral}, with the state $\sigma_{\text{Eve}}$ being 
\begin{equation}
    \sigma_{\text{Eve}}=\sum_{i_1=0 }^{d_{A_1}^2-1} \sum_{i_2=0 }^{d_{A_2}^2-1} \cdots \sum_{ i_n=0 }^{d_{A_n}^2-1}\left(\bigotimes_{j=1}^n U_{i_j}^*\right)\tr_{A_1A_2\cdots A_n}\left[\Xi_{i_1i_2\cdots i_n}\times \rho^{A_1A_2\cdots A_n}\right]\left(\bigotimes_{j=1}^n U_{i_j}^T\right),
\end{equation}
$\sigma_{\text{Eve}}$ can be prepared from $\rho$ by separable operations, and therefore $\sigma_{\text{Eve}}^T$ can also be prepared by separable operations. In the resource theory of entanglement, separable operations are all resource non-generating operations. Hence, Theorem \ref{MDIinGeneral}
still holds even if Eve is able to perform separable measurements. Since LOCC measurements are strictly weaker than separable measurements, we can safely conclude that in entanglement characterization tasks, classical communication can be allowed.

\section{Proof of Lemma \ref{POVMBound}}
\label{pfPOVMBound}

We consider the following process:
\begin{enumerate}
    \item Prepare a maximally entangled state $\Phi_+^{B_jA'_j}$ acting on $\mathcal{H}^{B_j}\otimes\mathcal{H}^{A'_j}$. Here, ${B_j}$ has the same dimension as $A_j$ and $A'_j$.
    \item Perform the POVM measurements $\Pi^{A_jA'_j}$ on $\mathcal{H}^{A_j}\otimes\mathcal{H}^{A'_j}$.
    \item Trace out the systems $A_1A_2\cdots A_n$ and $A'_1A'_2\cdots A'_n$. 
\end{enumerate}
After the entire process, we obtain a post-measurement state $\sigma_{i_1i_2\cdots i_n}$ with probability $p_{i_1i_2\cdots i_n}$:
\begin{equation}
    \sigma_{i_1i_2\cdots i_n}=\frac{\tilde{\rho}_{i_1i_2\cdots i_n}^T}{\tr\left(\tilde{\rho}_{i_1i_2\cdots i_n}\right)},\quad p_{i_1i_2\cdots i_n}=\frac{\tr\left(\tilde{\rho}_{i_1i_2\cdots i_n}\right)}{\Omega}.
\end{equation}

The whole procedure only consists of LOSR operations. Therefore, the following relation holds:
\begin{equation}
    \begin{aligned}
        \mathcal{E}({\rho})&\ge\sum_{i_1\in\mathcal{X}_1}\sum_{i_2\in\mathcal{X}_2}\cdots\sum_{i_n\in\mathcal{X}_n}p_{i_1i_2\cdots i_n}\mathcal{E}(\sigma_{i_1i_2\cdots i_n})\\
        &=\frac{1}{\Omega}\sum_{i_1\in\mathcal{X}_1}\sum_{i_2\in\mathcal{X}_2}\cdots\sum_{i_n\in\mathcal{X}_n}\mathcal{E}(\tilde{\rho}^T_{i_1i_2\cdots i_n})\\
        &=\frac{1}{\Omega}\sum_{i_1i_2\cdots i_n}\mathcal{E}(\tilde{\rho}_{i_1i_2\cdots i_n}),
    \end{aligned}
\end{equation}

\bibliographystyle{apsrev4-1}
\bibliography{MDIEW}

\end{document}